\documentclass[12pt]{article}

\usepackage{cite}
\usepackage{amsmath}
\usepackage{amssymb}
\usepackage{bm}
\usepackage{color}
\usepackage{graphicx}
\usepackage{multirow}
\usepackage{booktabs}
\numberwithin{equation}{section}

\allowdisplaybreaks

\DeclareMathOperator{\Det}{Det}

\newcounter{aff}

\begin{document}
\begin{titlepage}
\bigskip

\begin{flushright}
{\footnotesize OCU-PHYS 630}
\end{flushright}
\begin{center}
{\large\bf
Factorized Quantum Curves and Minuscule Vertices\\[6pt]
in 3D Duality Cascades with FI Parameters
}\\
\bigskip\bigskip
{\large
Sanefumi Moriyama\footnote{\tt moriyama@omu.ac.jp}
}\\
\bigskip
${}^{*}$\,{\it Department of Physics, Graduate School of Science,}\\
{\it Osaka Metropolitan University, Sumiyoshi-ku, Osaka 558-8585, Japan}
\end{center}

\begin{abstract}
In the study of duality cascades in three-dimensional gauge theories without FI parameters, an important role is played by a fundamental domain whose vertices correspond to brane configurations with vanishing relative ranks.
Through the Fermi gas formalism, such brane configurations are known to be represented by factorized quantum curves.
In this paper, we show that this factorized description extends naturally to quantum curves associated with del Pezzo geometries possessing exceptional Weyl-group symmetries in the presence of FI parameters.
We find that the vertices of the fundamental domain corresponding to the weights of minuscule representations are realized as factorized quantum curves built from canonical operators interpreted as 5-branes dressed with FI parameters.
This reveals an unexpected relation between factorized quantum curves and minuscule representations, providing a physical realization of these vertices in terms of ``extremal'' brane configurations.
\end{abstract}

\end{titlepage}

\tableofcontents

\section{Introduction}

Interestingly, duality transformations in supersymmetric gauge theories can be interpreted as brane exchanges in the corresponding brane configurations \cite{HW}, known as Hanany-Witten (HW) transitions.
When the brane configuration is compactified on a circle, one can successively apply HW transitions to change the ranks of gauge groups in three dimensions \cite{HW,A,HK}, similar to the well-known duality cascades in four dimensions \cite{KS}. 
The crucial difference is that three-dimensional dualities are IR dualities, meaning that dual UV theories flow to the same IR fixed point.
Hence, unlike in four dimensions, duality cascades are not dynamical along the renormalization-group flow.
Nevertheless, it is natural to seek the most economical dual description, where duality cascades should terminate.
We are then led to two natural questions: starting from an arbitrary brane configuration, do duality cascades always terminate after finitely many steps, and is the endpoint uniquely determined by the initial configuration, independent of the sequence of duality transformations?

This brane configuration on a circle becomes particularly interesting in connection with the proposal for the worldvolume theory of M2-branes \cite{ABJM}.
At the advent of M-theory, it was conjectured through the AdS/CFT correspondence \cite{KT} that the degrees of freedom of $N$ M2-branes scale as $N^{\frac{3}{2}}$ in the large $N$ limit.
This behavior was later reproduced in the ABJM theory from the partition function on $S^3$.
The ABJM theory \cite{ABJM,HLLLP2,ABJ} is the ${\cal N}=6$ supersymmetric Chern-Simons theory with gauge group $\text{U}(N_1)_{-k}\times\text{U}(N_2)_{k}$ (where the subscripts denote the Chern-Simons levels) and two pairs of bifundamental matter fields.
This theory is realized by compactifying on a circle the same brane configuration introduced in \cite{HW}, consisting of an NS5-brane, a $(1,k)$5-brane and D3-branes stretched between them.
After performing T-duality along the circle and lifting to M-theory, the brane configuration becomes one with $N=\min(N_1,N_2)$ M2-branes and $M=|N_2-N_1|$ fractional M2-branes on the background ${\mathbb C}^4/{\mathbb Z}_k$.
Thus, the overall rank $N$ and the relative rank $M$ play distinct roles in the ABJM theory.

Using the localization technique \cite{KWY}, the infinite-dimensional path integral defining the partition function on $S^3$ reduces to a finite-dimensional matrix integration.
Evaluating this matrix model for $\text{U}(N)_{-k}\times\text{U}(N)_{k}$ reproduces the expected $N^{\frac{3}{2}}$ behavior \cite{DMP1}.
Furthermore, all perturbative corrections can be resummed into an Airy function \cite{FHM}\footnote{For recent developments suggesting a broader appearance of Airy-function structures in superconformal indices and related observables, see \cite{Hri,BCHR,BT,BDHRZ,CDMMS}.} and nonperturbative instanton corrections were also investigated \cite{DMP1,DMP2}.
This structure led to the proposal of the Fermi gas formalism \cite{MP}, which not only reproduces the Airy function result in a simple way but also systematically clarifies the structure of instanton corrections.
Eventually, full expressions for the partition function, including worldsheet and membrane instanton effects, were obtained \cite{HMO2,HMO3,HMMO}, establishing the Fermi gas formalism as a powerful framework for studying M2-branes.

In the Fermi gas formalism \cite{MP}, the grand canonical partition function (obtained by dualizing the overall rank $N$) is given by the Fredholm determinant of a spectral operator taking the form of the curve $\mathbb{P}^1\times\mathbb{P}^1$ (whose coefficients encode the relative rank $M$).
Thus, the spectral operator is naturally regarded as the quantization of an algebraic curve \cite{MiMo,ACDKV}.
More generally, whenever the grand canonical partition function is described by a quantum curve of this type, the large-$N$ partition function universally exhibits the Airy-function behavior characteristic of M2-branes.
This suggests that quantum curves provide a natural framework for describing M2-branes, with the underlying brane configurations encoded in their geometry\footnote{For an interesting reinterpretation of quantum curves directly in terms of 5-brane configurations, see \cite{Kub}.}.

The ABJM theory corresponds to the simplest quantum curve $\mathbb{P}^1\times\mathbb{P}^1$, while its generalizations are associated with del Pezzo geometries enjoying exceptional Weyl-group symmetries $W(E_n)$.
These symmetries $W(E_n)$ naturally extend to the affine symmetries underlying $q$-Painlev\'e equations.
Indeed, the grand canonical partition functions of the ABJM theory and its generalization were found to satisfy the $q$-Painlev\'e III and VI equations respectively \cite{BGT,BGKNT,MN9,MNA}\footnote{See also \cite{GHM,NosM} for remarkable bilinear relations satisfied by the partition functions.}.

Returning to our original question concerning duality cascades, it is natural to expect that they also admit a description in terms of quantum curves.
Indeed, some HW transitions are identified with Weyl reflections of $W(E_n)$.
Furthermore, consistently with this picture, HW transitions change only the relative ranks (collectively denoted by ${\bm M}$), while leaving the overall rank $N$ invariant.

From the viewpoint of brane configurations and quantum curves, the questions of finiteness and uniqueness of duality cascades were studied in \cite{FMMN,FMS,MO} in relation to the theory of parallelotopes.
Each step of duality cascades is regarded as a discrete translation and the fundamental domain is defined as the region in the parameter space of relative ranks whose points no longer admit further cascades.
The questions of finiteness and uniqueness can then be reformulated as follows:
starting from an arbitrary point in the parameter space of relative ranks, does one always reach the fundamental domain after finitely many discrete translations, and is the resulting endpoint uniquely determined by the starting point?
Conversely, and equivalently, one may ask whether the fundamental domain tiles the entire space of relative ranks under these translations without overlaps or gaps.
Concisely speaking, the question is whether the polytope defining the fundamental domain is a parallelotope that tiles the whole space.

The problem of space tilings is a classical subject in mathematics \cite{G,Z,CS}.
Important classical results include the criterion determining when a zonotope, defined as the Minkowski sum of line segments, is a parallelotope \cite{She,McM}.
The Voronoi polytope of a lattice, consisting of all points at least as close to the origin as to any other lattice point, is also known to be a parallelotope.

Correspondingly, duality cascades in three dimensions have so far been studied mainly in two classes of setups.
First, for circular D3-brane configurations with arbitrary perpendicular 5-branes and vanishing Fayet--Iliopoulos (FI) parameters, the questions of finiteness and uniqueness were investigated in \cite{FMS,MO}.
In \cite{FMS}, three equivalent descriptions of the fundamental domain were introduced, following the terminology of polytope theory \cite{Z}.
The ${\cal H}$-description realizes the fundamental domain as the intersection of half-spaces determined by the condition that no further duality transformations are possible.
The ${\cal V}$-description identifies it as the convex hull of brane configurations with vanishing relative ranks, which become the vertices of the resulting polytope.
The ${\cal Z}$-description is based on the S-rule \cite{HW}, which bounds the number of D3-branes stretched between two 5-branes by the absolute value of the determinant of their charges.
The relative ranks are then parameterized by bounded nonnegative numbers associated with pairs of 5-branes, so that the fundamental domain becomes a zonotope.
These three descriptions were shown to be equivalent in \cite{FMS}.
Applying the classical criterion for zonotopes \cite{She,McM}, it was further proved that the fundamental domain is a parallelotope and hence tiles the entire space of relative ranks under discrete translations.
Consequently, under this setup, duality cascades always terminate after finitely many steps and their endpoints are uniquely determined.
By interpreting the circular quiver as the affine Dynkin quiver $\widehat A_n$, extensions to more general affine Dynkin quivers were also considered in \cite{MO}.

From the viewpoint of dimensions, the ${\cal H}$-description concerns codimension-one facets, whereas the ${\cal V}$-description involves zero-dimensional vertices and the ${\cal Z}$-description concerns one-dimensional edges.
Apparently, there is a substantial gap between the higher-dimensional ${\cal H}$-description and the lower-dimensional ${\cal V}$- and ${\cal Z}$-descriptions.

Second, duality cascades were studied from the viewpoint of quantum curves associated with del Pezzo geometries possessing exceptional Weyl-group symmetries $W(E_n)$ in \cite{FMMN}, rather than through the brane configurations themselves.
These quantum curves are realized as Laurent polynomials of the canonical operators $\widehat{Q}=e^{\widehat q}$ and $\widehat{P}=e^{\widehat p}$ satisfying $[\widehat q,\widehat p]=i\hbar$ with $\hbar=2\pi k$.
The corresponding Newton polygons are known to be triangular or rectangular \cite{Has,Tak,Mor,MY}, possibly with degeneracies in the asymptotic values \cite{BBT,KY}.

Although the complete interpretation of these quantum curves in terms of brane configurations is still lacking, several suggestive observations are known.
In the absence of FI parameters, the Fermi gas formalism \cite{MP} implies that, for brane configurations consisting only of NS5-branes and $(1,k)$5-branes with vanishing relative ranks, the quantum curve is constructed as a product of the two canonical operators corresponding to the two types of 5-branes,
\begin{align}
\widehat{\cal Q}=\widehat Q^{\frac{1}{2}}+\widehat Q^{-\frac{1}{2}},\quad
\widehat{\cal P}=\widehat P^{\frac{1}{2}}+\widehat P^{-\frac{1}{2}},
\label{QPhalf}
\end{align}
ordered in the reverse order of the original 5-brane configuration due to the inverse transformations in $\widehat{\cal Q}^{-1}=\big[2\cosh\frac{\widehat q}{2}\big]^{-1}$ and $\widehat{\cal P}^{-1}=\big[2\cosh\frac{\widehat p}{2}\big]^{-1}$.
After expanding the product and applying the Baker-Campbell-Hausdorff (BCH) formula, we find that these quantum curves consist of terms whose powers of $\widehat{Q}$ and $\widehat{P}$ each lie in fixed ranges.
This directly suggests a close relation between brane configurations and the rectangular realizations of quantum curves.
In particular, the numbers of NS5-branes and $(1,k)$5-branes determine the extent of the Newton polygon in the $\widehat Q$ and $\widehat P$ directions, respectively.
Although the coefficients generally become complicated after introducing relative ranks, the rectangular shape of the Newton polygon remains unchanged and quantum curves at the vertices of the fundamental domain are completely factorized into products of $\widehat{\cal Q}$ and $\widehat{\cal P}$.
Upon introducing FI parameters, the canonical operators are shifted by constants \cite{MP,Nos}.

The fundamental domain for these gauge theories associated with del Pezzo geometries was studied from the perspective of the exceptional Weyl group $W(E_n)$.
Starting from a single inequality imposed by the termination condition for duality cascades, the full set of inequalities is generated by the action of $W(E_n)$, yielding the ${\cal H}$-description of the fundamental domain.
The resulting domain is identified with the affine Weyl chamber (alcove), which is also the Voronoi polytope of the corresponding $E_n$ root lattice.
Although the vertices are obtained directly from the ${\cal H}$-description, their natural interpretation in the ${\cal V}$-description has so far remained unclear.
Given that the affine Weyl chamber is generally not a zonotope, it is particularly desirable to understand the lower-dimensional ${\cal V}$-description in terms of vertices.
Since the vertices correspond to factorized quantum curves without FI parameters, it is natural to expect that they continue to admit factorized realizations in the presence of FI parameters.

In this paper, we propose that factorized quantum curves constructed from
\begin{align}
\widehat{\cal Q}^{\pm}=\widehat Q^{\pm\frac{1}{2}}
(\widehat Q^{\frac{1}{2}}+\widehat Q^{-\frac{1}{2}})
=1+\widehat Q^{\pm 1},\quad
\widehat{\cal P}^{\pm}=\widehat P^{\pm\frac{1}{2}}
(\widehat P^{\frac{1}{2}}+\widehat P^{-\frac{1}{2}})
=1+\widehat P^{\pm 1},
\label{QPpm}
\end{align}
play a crucial role in studying the vertices of the fundamental domain for quantum curves associated with del Pezzo geometries possessing exceptional Weyl groups $D_5$, $E_6$ and $E_7$.
Namely, quantum curves at a distinguished subset of vertices, forming the minuscule representations of the corresponding Weyl groups, can be constructed by suitably ordering the canonical operators $\widehat{\cal Q}^\pm$ and $\widehat{\cal P}^\pm$ so as to realize the required degeneracies.
We find that these factorized curves reproduce precisely the corresponding vertices obtained from the ${\cal H}$-description.
This establishes a direct correspondence between factorized quantum curves and the minuscule vertices of the fundamental domain, providing a concrete realization of part of the ${\cal V}$-description for quantum curves associated with del Pezzo geometries.

In the absence of FI parameters, the vertices correspond to brane configurations with vanishing relative ranks and are described by the operators $\widehat{\cal Q}$ and $\widehat{\cal P}$, representing an NS5-brane and a $(1,k)$5-brane.
The appearance of $\widehat{\cal Q}^{\pm}$ and $\widehat{\cal P}^{\pm}$ therefore suggests an interpretation in terms of ``extremal'' 5-branes dressed with FI parameters\footnote{
FI parameters and mass deformations are closely related through Fourier transformations.
In the Fermi gas formalism, where quantum curves are themselves constructed by Fourier transformations, the distinction is often blurred.
Roughly speaking, the deformations are often referred to as FI parameters when the Newton polygon is fixed and as mass deformations when it changes.
Since we are interested here in the elementary building blocks, we collectively refer to them as FI parameters.
We are grateful to T.~Nosaka for valuable discussions on this point.}.
From the viewpoint of $q$-Painlev\'e equations \cite{KNY}, these operators provide elementary building blocks for distinguished sectors of quantum curves.
It would be interesting to clarify the role of these operators in both brane configurations and Painlev\'e systems.

In the next section, we first clarify the relation between brane configurations and quantum curves and propose a formulation of the fundamental domain of duality cascades in terms of quantum curves as a natural generalization of the description in terms of brane configurations.
We then study the fundamental domain defined by inequalities generated by the Weyl group and propose that its vertices are described by factorized quantum curves constructed from the canonical operators $\widehat{\cal Q}^\pm$ and $\widehat{\cal P}^\pm$ in \eqref{QPpm}.
We begin with the nondegenerate $D_5$ curve and subsequently turn to the $E_6$ and $E_7$ cases with degeneracies.
Finally, we conclude with future directions.

\section{Brane configurations and quantum curves}

Duality cascades in three dimensions have been studied from two complementary perspectives:
brane configurations in \cite{FMS,MO} and quantum curves in \cite{FMMN}.
In \cite{FMS} duality cascades for brane configurations with arbitrary numbers of 5-branes carrying arbitrary 5-brane charges (but without FI parameters) were analyzed.
Meanwhile, \cite{FMMN} studied quantum curves with exceptional Weyl-group symmetries for specific configurations including FI parameters, whose underlying physical interpretation remains less transparent.
In this section, we clarify the relation between these two approaches in order to gain insight into the vertices of the fundamental domain introduced in \cite{FMMN}, which will play a central role in the subsequent sections.

\subsection{Brane configurations}

\begin{table}[!t]
\begin{center}
\begin{tabular}{c||c|c|c|c|c}
directions&012&6&3\quad 7&4\quad 8&5\quad 9\\\hline\hline
D3&$-$&$-$&&&\\
NS5&$-$&&$-$\;\;\;\;\;&$-$\;\;\;\;\;&$-$\;\;\;\;\;\\
D5&$-$&&\;\;\;\;\;$-$&\;\;\;\;\;$-$&\;\;\;\;\;$-$\\
$(p,q)$5&$-$&&$[3,7]_\theta$&$[4,8]_\theta$&$[5,9]_\theta$
\end{tabular}
\end{center}
\caption{Directions along which the various branes extend.
$[i,j]_\theta$ denotes the direction $i$ tilted to $j$ by an angle $\theta=\arctan q/p$.}
\label{directions}
\end{table}

Let us recapitulate duality cascades in terms of brane configurations in this subsection.
We first specify the setup of brane configurations in type IIB string theory \cite{HW,ABJM}.
We place D3-branes along a circle in the $6$-direction, together with perpendicular $(n+1)$ 5-branes carrying charges $(p_i,q_i)$ located at different positions along the $6$-direction and tilted by angles $\theta_i=\arctan q_i/p_i$ to preserve supersymmetries.
The directions along which the various branes extend are summarized in table \ref{directions}.
The numbers of D3-branes may vary among different intervals between adjacent 5-branes.
Although the brane configuration possesses a cyclic symmetry along the $6$-direction, it is more convenient, when discussing duality cascades, to cut the circle into a line segment, analogous to taking local patches in a manifold.
Following \cite{FMS}, we refer to the interval at which the circle is cut as the reference interval, denote its endpoints by brackets $\langle\cdots\rangle$ and call the number of D3-branes $N_0$ in this interval the reference rank.
We also denote the 5-branes by $\overset{i}{\bullet}$ and record the numbers of D3-branes in each interval.
According to the HW transition, the system enjoys the symmetry of exchanging 5-branes,
\begin{align}
\cdots K\overset{i}{\bullet}L\overset{j}{\bullet}M\cdots
=\cdots K\overset{j}{\bullet}K-L+M+|k_{ij}|\overset{i}{\bullet}M\cdots,
\label{HW}
\end{align}
where the level for the gauge group $k_{ij}$ is determined by the determinant of the adjacent 5-brane charges
\begin{align}
k_{ij}=-\det\begin{pmatrix}p_i&q_i\\p_j&q_j\end{pmatrix}=-(p_i,q_i)\times(p_j,q_j).
\end{align}

For example, a brane configuration with four 5-branes is denoted by $\langle N_0\overset{1}{\bullet}N_1\overset{2}{\bullet}N_2\overset{3}{\bullet}N_3\overset{4}{\bullet}\rangle=\langle N_0\overset{1}{\bullet}N_0+M_1\overset{2}{\bullet}N_0+M_2\overset{3}{\bullet}N_0+M_3\overset{4}{\bullet}\rangle$.
Since the overall rank does not affect duality transformations \eqref{HW}, we often abbreviate the brane configuration as $\langle\overset{1}{\bullet}M_1\overset{2}{\bullet}M_2\overset{3}{\bullet}M_3\overset{4}{\bullet}\rangle$.

\begin{figure}[!t]
\vspace{-12mm}
\centering\includegraphics[width=16cm]{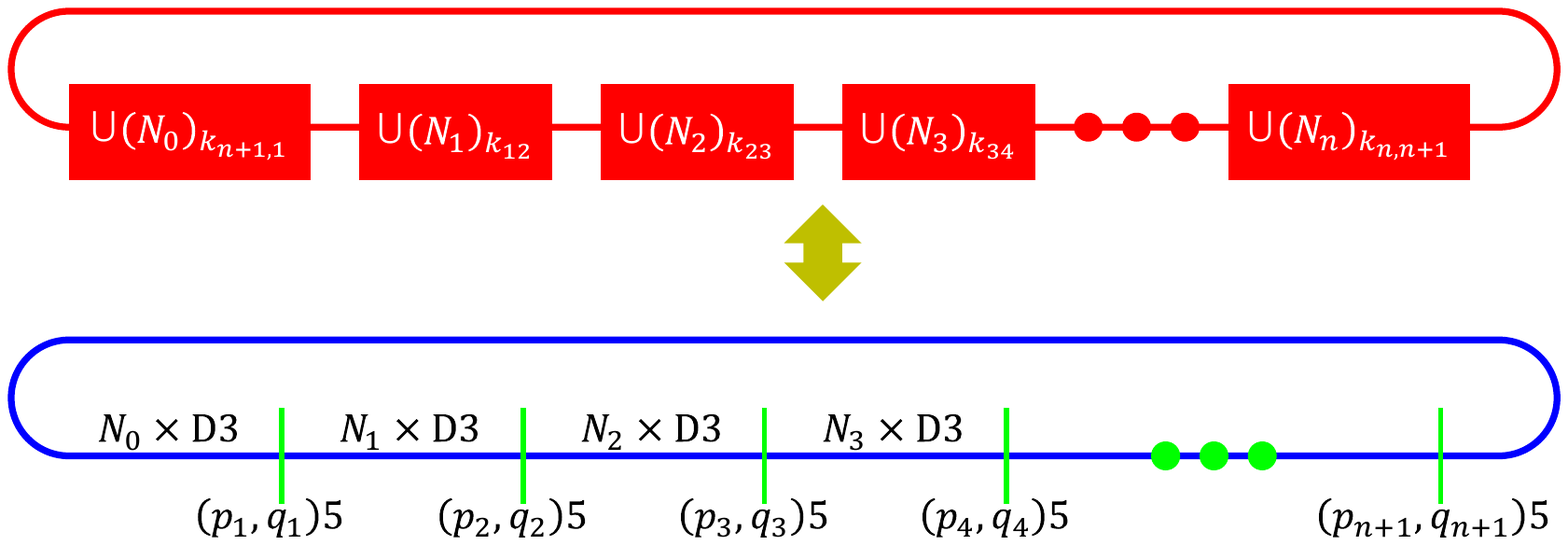}
\vspace{-40mm}
\caption{Correspondence between gauge theories and brane configurations.
A brane configuration of this type is often denoted by $\langle N_{0}\overset{1}{\bullet}N_1\overset{2}{\bullet}N_2\overset{3}{\bullet}\cdots\overset{n}{\bullet}N_n\overset{n+1}{\bullet}\rangle$.
The two descriptions can always be translated into each other when necessary.}
\label{gauge}
\end{figure}

The effective gauge theory described by this brane configuration is the ${\cal N}=3$ supersymmetric Chern-Simons theory with gauge group
\begin{align}
\text{U}(N_{0})_{k_{n+1,1}}\times\text{U}(N_1)_{k_{1,2}}\times\text{U}(N_2)_{k_{2,3}}\times\cdots\times\text{U}(N_{n})_{k_{n,n+1}},
\end{align}
and $(n+1)$ pairs of bifundamental matters connecting adjacent gauge group factors cyclically (see figure \ref{gauge}).
Although the gauge-theory description may be more familiar to some readers, we shall use the brane-configuration notation $\langle N_{0}\overset{1}{\bullet}N_1\overset{2}{\bullet}N_2\overset{3}{\bullet}\cdots\overset{n}{\bullet}N_n\overset{n+1}{\bullet}\rangle$ introduced above, since it provides a more economical description of the quantities relevant for duality cascades.
The corresponding gauge-theory description can always be recovered when needed.

We adopt the following working hypothesis for duality cascades \cite{FMMN,FMS}:
\begin{enumerate}
\item Apply the HW transitions arbitrarily, provided that the reference interval is not crossed.
\item Change the reference interval by cyclic rotations whenever a rank lower than the current reference rank appears.
\end{enumerate}
In this setup, it is natural to ask whether duality cascades always terminate after finitely many steps and whether the endpoint depends only on the initial brane configuration.

For example, let us consider the brane configuration ${\langle}7\overset{1}{\bullet}14\overset{2}{\bullet}18\overset{3}{\circ}{\rangle}$, where $\bullet$ and $\circ$ denote a $(1,k)$5-brane and an NS5-brane respectively with $k=3$ (see table \ref{exampleDC}).
Applying HW transitions \eqref{HW} without crossing the reference interval, we obtain ${\langle}7\overset{3}{\circ}2\overset{1}{\bullet}6\overset{2}{\bullet}{\rangle}$.
Since the rank $2$ is lower than the reference rank $7$, we change the reference interval by a cyclic rotation so that the reference rank becomes $2$, ${\langle}2\overset{1}{\bullet}6\overset{2}{\bullet}7\overset{3}{\circ}{\rangle}$.
Repeating this procedure, the configuration further reduces to ${\langle}2\overset{2}{\bullet}3\overset{3}{\circ}1\overset{1}{\bullet}{\rangle}$, where another reference change is required.
Eventually we arrive at ${\langle}1\overset{1}{\bullet}2\overset{2}{\bullet}3\overset{3}{\circ}{\rangle}$.
Since no rank lower than the reference rank appears under further HW transitions, the duality cascade terminates after finitely many steps.
We may instead perform the HW transitions in a different order starting from the original configuration.
The two choices shown in the table both lead to the same final configuration.
This suggests that the endpoint of a duality cascade is uniquely determined by the initial configuration.

\begin{table}[!t]
\begin{align*}
{\langle}7\overset{1}{\bullet}14\overset{2}{\bullet}18\overset{3}{\circ}{\rangle}
\begin{array}{c}
\rotatebox[origin=c]{30}{$=$}\\
\rotatebox[origin=c]{-30}{$=$}
\end{array}\!\!\!
\begin{array}{c}
{\langle}7\overset{3}{\circ}2\overset{1}{\bullet}6\overset{2}{\bullet}{\rangle}\\
{\langle}7\overset{2}{\bullet}11\overset{3}{\circ}3\overset{1}{\bullet}{\rangle}
\end{array}\!\!\!
\begin{array}{c}
\to\\\to
\end{array}\!\!\!
\begin{array}{c}
{\langle}2\overset{1}{\bullet}6\overset{2}{\bullet}7\overset{3}{\circ}{\rangle}\\
{\langle}3\overset{1}{\bullet}7\overset{2}{\bullet}11\overset{3}{\circ}{\rangle}
\end{array}\!\!\!
\begin{array}{c}
=\\=
\end{array}\!\!\!
\begin{array}{c}
{\langle}2\overset{2}{\bullet}3\overset{3}{\circ}1\overset{1}{\bullet}{\rangle}\\
{\langle}3\overset{3}{\circ}1\overset{1}{\bullet}2\overset{2}{\bullet}{\rangle}
\end{array}\!\!\!
\begin{array}{c}
\rotatebox[origin=c]{-30}{$\to$}\\
\rotatebox[origin=c]{30}{$\to$}
\end{array}
{\langle}1\overset{1}{\bullet}2\overset{2}{\bullet}3\overset{3}{\circ}{\rangle}
\end{align*}
\caption{An example of duality cascades.
$\bullet$ and $\circ$ denote a $(1,k)$5-brane and an NS5-brane respectively with $k=3$.
The example illustrates the notions of finiteness and uniqueness of duality cascades.}
\label{exampleDC}
\end{table}

By defining the fundamental domain of duality cascades as the region in the parameter space of relative ranks for which no further duality cascades occur, the questions of finiteness and uniqueness can be reformulated in terms of the geometrical properties of the fundamental domain \cite{FMMN}.
Since charge conservation implies that each step of duality cascades acts as a discrete translation in the parameter space, the finiteness and uniqueness questions can be restated as follows:
starting from an arbitrary point in the parameter space of relative ranks, does the point always reach the fundamental domain after finitely many discrete translations, and is the endpoint uniquely determined?
Conversely, and equivalently, one may ask whether the fundamental domain tiles the entire parameter space under discrete translations without overlaps or gaps.
In summary, the problem reduces to determining whether the fundamental domain is a parallelotope.

In \cite{FMS,MO} these questions were answered using brane configurations.
Although the arguments apply to an arbitrary number $(n+1)$ of 5-branes carrying arbitrary 5-brane charges, explicit examples with $n=2$ and $n=3$ were presented in \cite{MO} and \cite{FMS}, respectively.
Here we revisit the example of $n=3$, using a slightly different parameterization of relative ranks, $\langle\overset{1}{\bullet}M_1\overset{2}{\bullet}M_2\overset{3}{\bullet}M_3\overset{4}{\bullet}\rangle$.

\begin{figure}[!t]
\vspace{-12mm}
\centering\includegraphics[width=16cm]{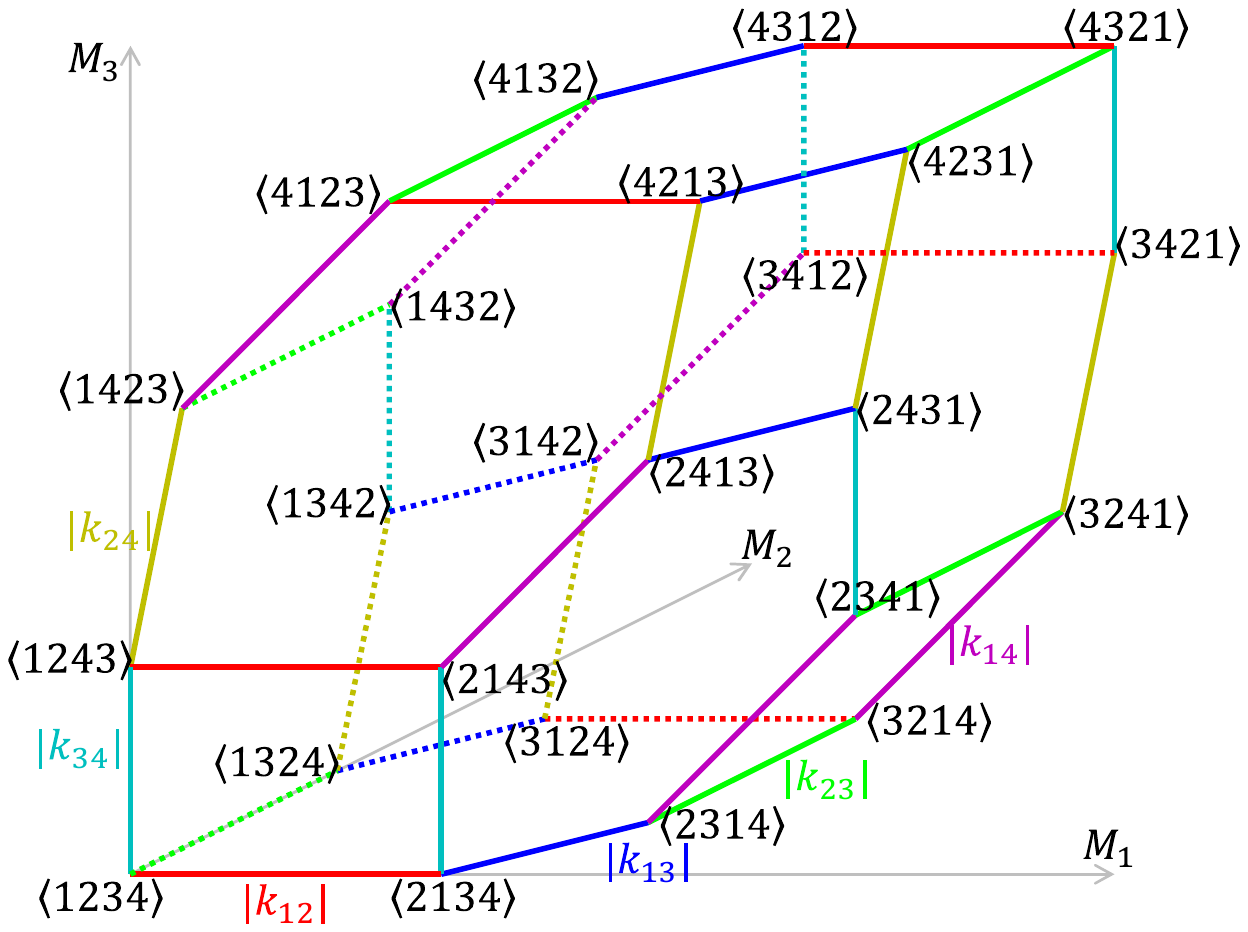}
\vspace{-16mm}
\caption{Three descriptions of the fundamental domain for a circular brane system with four 5-branes carrying arbitrary charges (without FI parameters).
The ${\cal H}$-description is defined by the inequalities determining the facets, such as those in \eqref{H4ineq}, while the ${\cal V}$-description is given by brane configurations with vanishing relative ranks in different orderings, such as \eqref{V4} and \eqref{3142}.
In the figure, the symbols $\bullet$ representing 5-branes are omitted for simplicity, so that $\langle 3142\rangle$ stands for ${\langle}{\overset{3}{\bullet}}{\overset{1}{\bullet}}{\overset{4}{\bullet}}{\overset{2}{\bullet}}{\rangle}$.
The ${\cal Z}$-description realizes the fundamental domain as the zonotope generated by the six line segments \eqref{zone}, shown in different colors.
The resulting polytope is affinely equivalent to the truncated octahedron.
Comparing the ${\cal V}$- and ${\cal Z}$-descriptions, one finds that each vertex is obtained from the standard ordering ${\langle}1234{\rangle}$ by a sequence of adjacent 5-brane exchanges, corresponding precisely to the generators of the zonotope.
This mirrors the decomposition of permutations into transpositions.
For example, the vertex $\langle 3142\rangle$ can be reached from $\langle 1234\rangle$ through the two reduced decompositions $(24)(13)(23)$ and $(13)(24)(23)$, corresponding to two edge paths of the polytope.
Equivalently, its relative ranks are obtained from those of $\langle 4321\rangle$ by setting to zero the remaining generators, proportional to $|k_{12}|$, $|k_{14}|$ and $|k_{34}|$.
}
\label{four5}
\end{figure}

We begin with three descriptions of the fundamental domain.
The first, called the ${\cal H}$-description, is based directly on the definition of duality cascades.
Requiring that duality cascades terminate implies that no ranks lower than the reference rank can appear.
This condition gives rise to a collection of linear inequalities, and the intersection of the corresponding half-spaces defines the fundamental domain.
The second, the ${\cal V}$-description, assumes that brane configurations with vanishing relative ranks serve as the vertices of the fundamental domain and takes their convex hull.
The third is the ${\cal Z}$-description, which utilizes the supersymmetry condition known as the S-rule, stating that the number of D3-branes stretched between two 5-branes $i$ and $j$ is nonnegative and bounded above by the level $|k_{ij}|$.
This implies that the relative ranks are obtained as projections of the direct product of intervals $[0,|k_{ij}|]$, namely a zonotope.
The inclusion relation ${\cal V}\subset{\cal Z}\subset{\cal H}$ follows immediately from the definitions.
Furthermore, \cite{FMS} proved the equivalence ${\cal V}={\cal Z}={\cal H}$.

For example, in the case of four 5-branes, the HW transitions \eqref{HW} allows one to reverse the ordering of 5-branes as
\begin{align}
&\langle\overset{1}{\bullet}M_1\overset{2}{\bullet}M_2\overset{3}{\bullet}M_3\overset{4}{\bullet}\rangle\nonumber\\
&=\langle\overset{4}{\bullet}|k_{14}|+|k_{24}|+|k_{34}|-M_3\overset{3}{\bullet}|k_{13}|+|k_{14}|+|k_{23}|+|k_{24}|-M_2\overset{2}{\bullet}|k_{12}|+|k_{13}|+|k_{14}|-M_1\overset{1}{\bullet}\rangle.
\label{H4}
\end{align}
Requiring that no rank lower than the reference rank appears in the ${\cal H}$-description then yields the inequalities
\begin{align}
&0\le M_1\le|k_{12}|+|k_{13}|+|k_{14}|,\nonumber\\
&0\le M_2\le|k_{13}|+|k_{14}|+|k_{23}|+|k_{24}|,\nonumber\\
&0\le M_3\le|k_{14}|+|k_{24}|+|k_{34}|,
\label{H4ineq}
\end{align}
together with eight additional inequalities.
Altogether, the fundamental domain becomes a truncated octahedron (after an appropriate affine transformation), as depicted in figure \ref{four5}.

In the ${\cal V}$-description, one considers all possible orderings of 5-branes with vanishing relative ranks.
To represent these configurations as points in the parameter space of relative ranks ${\bm M}$ and take their convex hull, we fix a standard ordering of 5-branes and bring each configuration to this ordering by HW transitions \eqref{HW}.
For example, starting from the reverse ordering ${\langle}{\overset{4}{\bullet}}{\overset{3}{\bullet}}{\overset{2}{\bullet}}{\overset{1}{\bullet}}{\rangle}$, one can use HW transitions to bring the configuration to the standard ordering ${\langle}{\overset{1}{\bullet}}{\overset{2}{\bullet}}{\overset{3}{\bullet}}{\overset{4}{\bullet}}{\rangle}$ and obtain
\begin{align}
{\langle}{\overset{4}{\bullet}}{\overset{3}{\bullet}}{\overset{2}{\bullet}}{\overset{1}{\bullet}}{\rangle}
=\langle\overset{1}{\bullet}|k_{12}|+|k_{13}|+|k_{14}|\overset{2}{\bullet}|k_{13}|+|k_{14}|+|k_{23}|+|k_{24}|\overset{3}{\bullet}|k_{14}|+|k_{24}|+|k_{34}|\overset{4}{\bullet}\rangle,
\label{V4}
\end{align}
from which one reads off the vertex ${\bm M}=(M_1,M_2,M_3)=(|k_{12}|+|k_{13}|+|k_{14}|,|k_{13}|+|k_{14}|+|k_{23}|+|k_{24}|,|k_{14}|+|k_{24}|+|k_{34}|)$.
Similarly, from
\begin{align}
{\langle}{\overset{3}{\bullet}}{\overset{1}{\bullet}}{\overset{4}{\bullet}}{\overset{2}{\bullet}}{\rangle}
=\langle\overset{1}{\bullet}|k_{13}|\overset{2}{\bullet}|k_{13}|+|k_{23}|+|k_{24}|\overset{3}{\bullet}|k_{24}|\overset{4}{\bullet}\rangle,
\label{3142}
\end{align}
we obtain ${\bm M}=(|k_{13}|,|k_{13}|+|k_{23}|+|k_{24}|,|k_{24}|)$.
In total, one obtains $4!=24$ vertices corresponding to the various orderings of the four 5-branes and takes their convex hull.

Finally, the ${\cal Z}$-description identifies the fundamental domain as the zonotope generated by the six line segments
\begin{align}
|k_{12}|(1,0,0),\;|k_{13}|(1,1,0),\;|k_{14}|(1,1,1),\;|k_{23}|(0,1,0),\;|k_{24}|(0,1,1),\;|k_{34}|(0,0,1),
\label{zone}
\end{align}
associated with each pair of 5-branes through the S-rule.
Note that brane configurations at the vertices can be reached from the configuration in the standard ordering ${\langle}{\overset{1}{\bullet}}{\overset{2}{\bullet}}{\overset{3}{\bullet}}{\overset{4}{\bullet}}{\rangle}$ with sums of the vectors \eqref{zone} depending on which pairs of 5-branes are reversed.
For ${\langle}{\overset{4}{\bullet}}{\overset{3}{\bullet}}{\overset{2}{\bullet}}{\overset{1}{\bullet}}{\rangle}$, all pairs of 5-branes are reversed and all the vectors in \eqref{zone} are used, while for ${\langle}{\overset{3}{\bullet}}{\overset{1}{\bullet}}{\overset{4}{\bullet}}{\overset{2}{\bullet}}{\rangle}$, only those proportional to $|k_{13}|$, $|k_{23}|$ and $|k_{24}|$ are used.
Namely, the relative ranks ${\bm M}$ for a given vertex can be obtained from those of the completely reversed ordering by setting to zero the levels associated with pairs of 5-branes whose ordering is not reversed.
In the example of ${\langle}{\overset{3}{\bullet}}{\overset{1}{\bullet}}{\overset{4}{\bullet}}{\overset{2}{\bullet}}{\rangle}$, this amounts to setting $|k_{12}|$, $|k_{14}|$ and $|k_{34}|$ to zero (compare \eqref{V4} and \eqref{3142}).

After establishing the equivalence of the three descriptions, one can apply the parallelotope criterion for zonotopes \cite{She,McM} and conclude that the fundamental domain is indeed a parallelotope.
Consequently, under this setup, duality cascades always terminate after finitely many steps and the endpoint is uniquely determined.
In the above discussions of parallelotopes, apparently the lower-dimensional ${\cal Z}$-description plays a crucial role.

This argument for the questions of finiteness and uniqueness of duality cascades applies to general brane configurations with an arbitrary number of 5-branes carrying arbitrary charges.
However, it does not incorporate the additional field-theoretical parameters known as FI parameters.
On the other hand, duality cascades have also been studied from the viewpoint of quantum curves with FI parameters \cite{FMMN}.
In that setup, the fundamental domain is identified with an affine Weyl chamber, but little is known beyond this ${\cal H}$-description.
Since the affine Weyl chamber is generally not a zonotope, the powerful ${\cal Z}$-description available for brane configurations is absent.
It is therefore interesting to clarify the closely related ${\cal V}$-description.
Before turning to quantum curves associated with del Pezzo geometries with FI parameters in the next section, we first study an analogous setup without FI parameters from the viewpoint of quantum curves.

\subsection{Quantum curves}

In the previous subsection we reviewed duality cascades in brane configurations without FI parameters.
In the discussions, we encoded the information about 5-brane orderings into relative ranks by utilizing the HW transitions \eqref{HW} in \eqref{H4} and \eqref{V4}, \eqref{3142} for the ${\cal H}$- and ${\cal V}$-descriptions.
Before turning to quantum curves with FI parameters in the subsequent sections, we would first like to understand what characterizes the quantum curves at the vertices and how information about relative ranks appears in quantum curves.
For this purpose, we reformulate the discussions in brane configurations in terms of quantum curves.
In this subsection, we show that quantum curves at the vertices take completely factorized forms; that the BCH formula plays the same role for quantum curves as the HW transitions \eqref{HW} do for brane configurations; and that relative ranks appear as asymptotic values of the curve.

The grand canonical partition function for D3-branes in the presence of perpendicular $(p_i,q_i)$5-branes $(p_i>0)$ is obtained from the partition function by introducing the fugacity $\kappa$ conjugate to the overall rank $N$,
\begin{align}
\Xi_{{\bm k},{\bm M}}(\kappa)
=\sum_{N=0}^\infty\kappa^NZ_{{\bm k},{\bm M}}(N).
\label{GPF}
\end{align}
In the following, we often suppress the subscripts denoting levels ${\bm k}$ and relative ranks ${\bm M}$, since our discussion will later be extended to quantum curves whose brane interpretations are not necessarily clear.
It was shown \cite{MP} that, without FI parameters, the grand canonical partition function is given, up to normalization, by the Fredholm determinant of a spectral operator
\begin{align}
\Xi(\kappa)\simeq\Det(1+\kappa/\widehat H),
\label{Fredholm}
\end{align}
with the spectral operator given by
\begin{align}
\widehat H=\prod_{i=1}^{n+1}\Big(2\cosh\frac{p_i\widehat p-q_i\widehat q}{2}\Big),
\end{align}
where the product is noncommutative, with $[\widehat q,\widehat p]=2\pi i$, and is taken in the reverse order of the 5-branes.
The reverse ordering originates from the inverse operators appearing in the Fermi gas formalism.
If we introduce the operator
\begin{align}
\widehat E_i=e^{p_i\widehat p-q_i\widehat q},
\end{align}
the spectral operator is rewritten as
\begin{align}
\widehat H=\prod_{i=1}^{n+1}(\widehat E_i^{\frac{1}{2}}+\widehat E_i^{-\frac{1}{2}}).
\label{factorize}
\end{align}

Although this product is noncommutative, we can combine exponentials using the BCH formula
\begin{align}
e^{\epsilon_i\frac{p_i\widehat p-q_i\widehat q}{2}}e^{\epsilon_j\frac{p_j\widehat p-q_j\widehat q}{2}}
=e^{\epsilon_i\frac{p_i\widehat p-q_i\widehat q}{2}+\epsilon_j\frac{p_j\widehat p-q_j\widehat q}{2}-\epsilon_i\epsilon_j\frac{\pi i}{4}k_{ij}}
\end{align}
where $\epsilon_{l(=i,j)}=\pm 1$ denote independent sign choices for $\widehat E_{l}^{\frac{\epsilon_{l}}{2}}$ in $\widehat E_{l}^{\frac{1}{2}}+\widehat E_{l}^{-\frac{1}{2}}$.
Note that, as in \cite{KMZ}, we analytically continue the levels ${\bm k}=(k_{ij})$ and the ranks ${\bm M}$ to the covering space and do not impose the integrality conditions.
This allows the noncommutative phase factors arising from the BCH formula to remain nontrivial.
We often abbreviate this formula as
\begin{align}
\widehat E_i^{\frac{\epsilon_i}{2}}\widehat E_j^{\frac{\epsilon_j}{2}}
=\big[\widehat E_i^{\frac{\epsilon_i}{2}}\widehat E_j^{\frac{\epsilon_j}{2}}\big]e^{-\frac{\pi i}{4}\epsilon_i\epsilon_jk_{ij}},
\label{BCH}
\end{align}
where the parentheses $\big[\cdots\big]$ denote the Weyl ordering.
We can then combine the various terms appearing in the expansion.

\begin{figure}[!t]
\vspace{-12mm}
\centering\includegraphics[width=16cm]{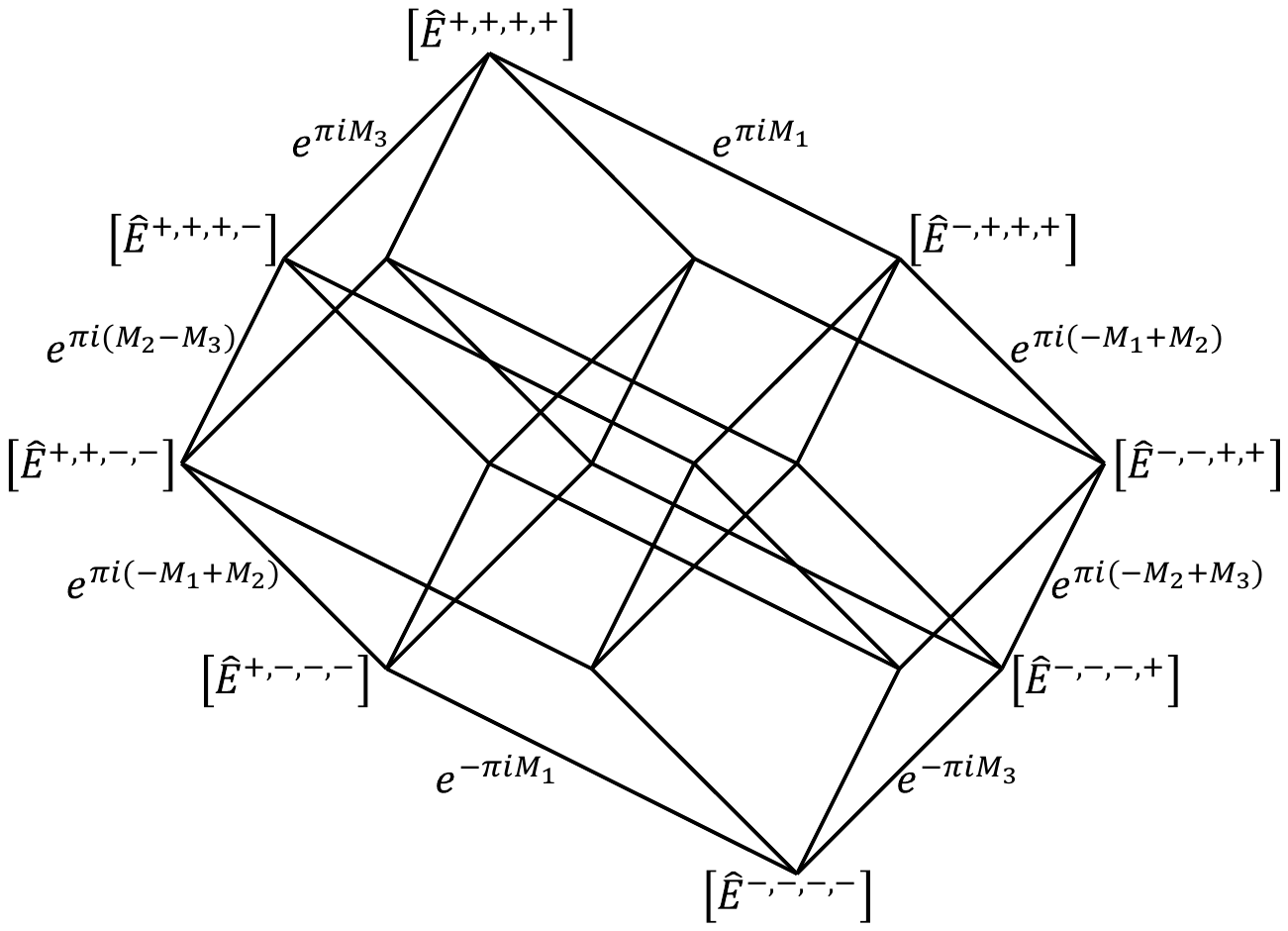}
\vspace{-16mm}
\caption{Newton polygon for quantum curves corresponding to brane configurations with four 5-branes.
Relative coefficients along the boundary of the Newton polygon (corresponding to asymptotic values) are expressed in terms of the relative ranks ${\bm M}$.}
\label{boundaryNP}
\end{figure}

As in the ${\cal V}$-description, where a given ordering is compared with the standard ordering through HW transitions, we compare the coefficients of the factorized quantum curve with those of the quantum curve in the standard ordering.
For example, let us study the quantum curve
\begin{align}
\widehat H_{1234}=
(\widehat E_1^{\frac{1}{2}}+\widehat E_1^{-\frac{1}{2}})
(\widehat E_2^{\frac{1}{2}}+\widehat E_2^{-\frac{1}{2}})
(\widehat E_3^{\frac{1}{2}}+\widehat E_3^{-\frac{1}{2}})
(\widehat E_4^{\frac{1}{2}}+\widehat E_4^{-\frac{1}{2}}),
\label{E1234}
\end{align}
corresponding to the brane configuration ${\langle}{\overset{4}{\bullet}}{\overset{3}{\bullet}}{\overset{2}{\bullet}}{\overset{1}{\bullet}}{\rangle}$.
In the ${\cal V}$-description \eqref{V4}, we bring this brane configuration ${\langle}{\overset{4}{\bullet}}{\overset{3}{\bullet}}{\overset{2}{\bullet}}{\overset{1}{\bullet}}{\rangle}$ into the standard ordering ${\langle}{\overset{1}{\bullet}}{\overset{2}{\bullet}}{\overset{3}{\bullet}}{\overset{4}{\bullet}}{\rangle}$ with HW transitions.
Here we expand the factors \eqref{E1234} using the BCH formula \eqref{BCH},
\begin{align}
\widehat E_1^{\frac{\epsilon_1}{2}}\widehat E_2^{\frac{\epsilon_2}{2}}\widehat E_3^{\frac{\epsilon_3}{2}}\widehat E_4^{\frac{\epsilon_4}{2}}
=c_{\epsilon_1,\epsilon_2,\epsilon_3,\epsilon_4}
\big[\widehat E^{\epsilon_1,\epsilon_2,\epsilon_3,\epsilon_4}\big],\quad
c_{\epsilon_1,\epsilon_2,\epsilon_3,\epsilon_4}=e^{-\frac{\pi i}{4}\sum_{i<j}\epsilon_i\epsilon_jk_{ij}},
\label{E1234BCH}
\end{align}
with the shorthand notation for the Weyl-ordered product
\begin{align}
\big[\widehat E^{\epsilon_1,\epsilon_2,\epsilon_3,\epsilon_4}\big]=\big[\widehat E_1^{\frac{\epsilon_1}{2}}\widehat E_2^{\frac{\epsilon_2}{2}}\widehat E_3^{\frac{\epsilon_3}{2}}\widehat E_4^{\frac{\epsilon_4}{2}}\big].
\end{align}
We then compare the results with those of $\widehat H_{4321}$ in the standard ordering of ${\langle}{\overset{1}{\bullet}}{\overset{2}{\bullet}}{\overset{3}{\bullet}}{\overset{4}{\bullet}}{\rangle}$.
Here we assume $q_i/p_i>q_j/p_j$ for $i<j$, so that $k_{ij}>0$.
Then the terms
\begin{equation}
\big[\widehat E^{+,+,+,+}\big],\quad
\begin{array}{c}
\big[\widehat E^{-,+,+,+}\big],\\
\vdots\\
\big[\widehat E^{+,+,+,-}\big],
\end{array}\quad
\begin{array}{c}
\big[\widehat E^{-,-,+,+}\big],\\
\vdots\\
\big[\widehat E^{+,+,-,-}\big],
\end{array}\quad
\begin{array}{c}
\big[\widehat E^{-,-,-,+}\big],\\
\vdots\\
\big[\widehat E^{+,-,-,-}\big],
\end{array}\quad
\big[\widehat E^{-,-,-,-}\big],
\end{equation}
appear on the boundary of the corresponding Newton polygon in figure \ref{boundaryNP}.

Adjacent boundary terms of the Newton polygon differ by the sign of a single $\epsilon_i$.
Hence, the relative coefficient between adjacent boundary terms is determined by the contribution associated with the flipped sign.
By regarding the spectral operator as a classical curve on the canonical phase space, these relative coefficients correspond to the asymptotic values of the curve.
For the example of $\widehat H_{1234}$ \eqref{E1234} corresponding to ${\langle}{\overset{4}{\bullet}}{\overset{3}{\bullet}}{\overset{2}{\bullet}}{\overset{1}{\bullet}}{\rangle}$, with the formula
\begin{align}
\frac{c_{-,\epsilon_2,\epsilon_3,\epsilon_4}}{c_{+,\epsilon_2,\epsilon_3,\epsilon_4}}
=e^{\frac{\pi i}{2}(\epsilon_2k_{12}+\epsilon_3k_{13}+\epsilon_4k_{14})},\quad
\frac{c_{\epsilon_1,-,\epsilon_3,\epsilon_4}}{c_{\epsilon_1,+,\epsilon_3,\epsilon_4}}
=e^{\frac{\pi i}{2}(\epsilon_1k_{12}+\epsilon_3k_{23}+\epsilon_4k_{24})},\quad
\cdots,
\end{align}
the coefficient of $\big[\widehat E^{-,+,+,+}\big]$ relative to $\big[\widehat E^{+,+,+,+}\big]$ is $e^{\frac{\pi i}{2}(k_{12}+k_{13}+k_{14})}$ since $\epsilon_1$ is flipped, while that of $\big[\widehat E^{+,+,+,-}\big]$ is $e^{\frac{\pi i}{2}(k_{14}+k_{24}+k_{34})}$ since $\epsilon_4$ is flipped.
Furthermore, the relative coefficient between $\big[\widehat E^{-,+,+,+}\big]$ and $\big[\widehat E^{-,-,+,+}\big]$ is $e^{\frac{\pi i}{2}(-k_{12}+k_{23}+k_{24})}$, while that between $\big[\widehat E^{+,+,+,-}\big]$ and $\big[\widehat E^{+,+,-,-}\big]$ is $e^{\frac{\pi i}{2}(k_{13}+k_{23}-k_{34})}$ and so on.
The relative coefficients along the boundary are summarized as
\begin{align}
r(k_{ij})=\left\{\begin{matrix}e^{\frac{\pi i}{2}(k_{12}+k_{13}+k_{14})},&e^{\frac{\pi i}{2}(-k_{12}+k_{23}+k_{24})},&e^{\frac{\pi i}{2}(-k_{13}-k_{23}+k_{34})},&e^{\frac{\pi i}{2}(-k_{14}-k_{24}-k_{34})}\\e^{\frac{\pi i}{2}(k_{14}+k_{24}+k_{34})},&e^{\frac{\pi i}{2}(k_{13}+k_{23}-k_{34})},&e^{\frac{\pi i}{2}(k_{12}-k_{23}-k_{24})},&e^{\frac{\pi i}{2}(-k_{12}-k_{13}-k_{14})}\end{matrix}\right\}.
\end{align}

We further compare these relative coefficients for $\widehat H_{1234}$ with those for the standard-ordering curve $\widehat H_{4321}$ corresponding to ${\langle}{\overset{1}{\bullet}}{\overset{2}{\bullet}}{\overset{3}{\bullet}}{\overset{4}{\bullet}}{\rangle}$.
Since $\widehat H_{4321}$ is completely reversed relative to $\widehat H_{1234}$, its relative coefficients are obtained by replacing $k_{ij}$ with $-k_{ij}$, namely $r(k_{ij})\to r(-k_{ij})$.
Consequently, taking the component-wise ratio gives $r(k_{ij})/r(-k_{ij})=r(2k_{ij})$ reflecting the doubling of the exponents.
We then find that the boundary coefficients can be expressed in terms of the relative ranks $(M_1,M_2,M_3)$ in \eqref{V4} as
\begin{align}
\left\{\begin{matrix}
e^{\pi iM_1},&
e^{\pi i(-M_1+M_2)},&
e^{\pi i(-M_2+M_3)},&
e^{-\pi iM_3}\\
e^{\pi iM_3},&
e^{\pi i(M_2-M_3)},&
e^{\pi i(M_1-M_2)},&
e^{-\pi iM_1}\end{matrix}\right\}.
\label{boundary4}
\end{align}

Although our first example $\widehat H_{1234}$ corresponds to the complete reversal of the standard ordering $\widehat H_{4321}$, the result \eqref{boundary4} in figure \ref{boundaryNP} remains valid for arbitrary orderings.
Indeed, since all pairs of 5-branes are reversed between $\widehat H_{1234}$ and $\widehat H_{4321}$, the relative coefficients in the latter are obtained from those in the former by replacing every $k_{ij}$ with $-k_{ij}$.
Hence, each contribution is doubled in the comparison, $k_{ij}-(-k_{ij})=2k_{ij}$.
For a general ordering, the corresponding quantum curve is obtained from $\widehat H_{1234}$ by reversing the signs of $k_{ij}$ only for those pairs whose relative ordering differs from that in $\widehat H_{1234}$.
Hence, when compared with the standard ordering, pairs whose ordering agrees with that of $\widehat H_{1234}$ again contribute $2k_{ij}$, while pairs whose ordering differs contribute zero, $(-k_{ij})-(-k_{ij})=0$, and therefore drop out.
This is equivalent to setting the corresponding $k_{ij}$ to zero in $r(2k_{ij})$.
For example, for $\widehat H_{2413}$, the signs for $k_{12}$, $k_{14}$ and $k_{34}$ are reversed relative to $\widehat H_{1234}$, since the corresponding pairs of 5-branes are reversed.
Consequently, these contributions cancel when compared with the standard ordering.
This is precisely the same mechanism as in the ${\cal Z}$-description, where only the line segments in \eqref{zone} associated with reversed pairs contribute to the relative ranks.
In the present language, this corresponds to setting the remaining $k_{ij}$ to zero in $r(2k_{ij})$.

Therefore, the information about relative ranks encoded in HW transitions is reproduced entirely by the coefficient differences of factorized quantum curves.
In this sense, one can study relative ranks using quantum curves instead of brane configurations.

More generally, for $(n+1)$ 5-branes, the relative coefficients of the boundary terms of the Newton polygon are given by
\begin{align}
\left\{\begin{matrix}
e^{\pi iM_1},&
e^{\pi i(-M_1+M_2)},&
\cdots,&
e^{\pi i(-M_{n-1}+M_n)},&
e^{-\pi iM_n}\\
e^{\pi iM_n},&
e^{\pi i(M_{n-1}-M_n)},&
\cdots,&
e^{\pi i(M_1-M_2)},&
e^{-\pi iM_1}\end{matrix}\right\}.
\label{boundary}
\end{align}
Note, however, that the coefficients associated with internal lattice points of the Newton polygon are generally more complicated and need not be expressed as linear functions of $k_{ij}$.

To summarize, information about brane configurations is encoded in the asymptotic values of the boundary terms of the Newton polygon.
We stress that, although the coefficients become complicated after applying the BCH formula, quantum curves at the vertices are special in that they are completely factorized as in \eqref{factorize}.
The picture we obtain is that, while quantum curves at generic points of the fundamental domain are described as sums of operator monomials with arbitrary coefficients, the same curves become factorized into products of noncommutative operators precisely at the vertices.

\begin{figure}[!t]
\centering
\includegraphics[width=12cm]{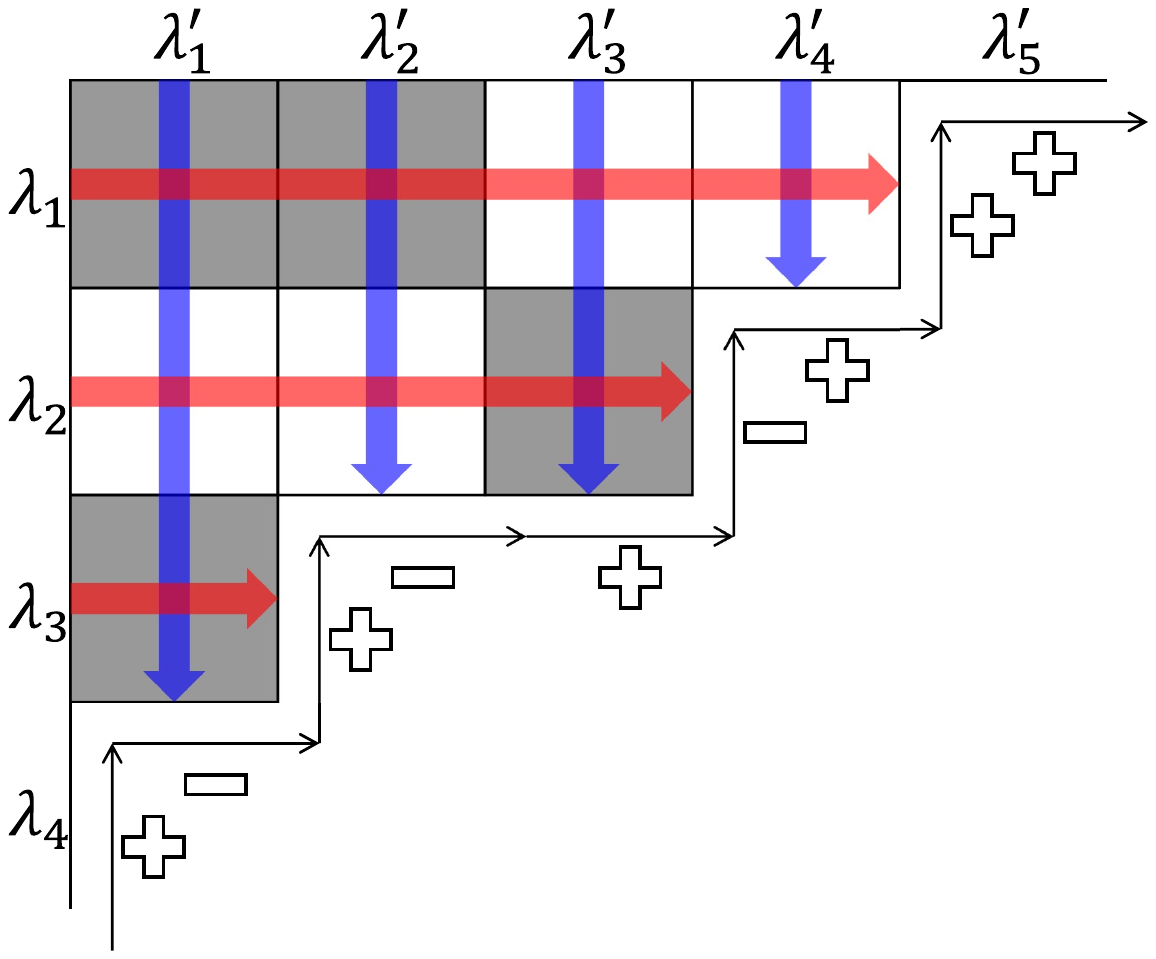}
\vspace{-12mm}
\caption{Young diagram $[\lambda_1,\lambda_2,\lambda_3,\lambda_4]=[4,3,1,0]$ corresponding to the operator ordering $\widehat{\cal P}\widehat{\cal Q}\widehat{\cal P}\widehat{\cal Q}^2\widehat{\cal P}\widehat{\cal Q}\widehat{\cal P}\widehat{\cal Q}$.
We associate the operator ordering with a Young diagram by identifying $\widehat{\cal P}$ with a step upward and $\widehat{\cal Q}$ with a step to the right when reading the spectral operator $\widehat{\cal H}$ from left to right.
We may independently choose either sign in $\widehat{\cal Q}=\widehat Q^{\frac{1}{2}}+\widehat Q^{-\frac{1}{2}}$ and $\widehat{\cal P}=\widehat P^{\frac{1}{2}}+\widehat P^{-\frac{1}{2}}$.
The figure illustrates the sign assignment corresponding to $\widehat{P}^{\frac{1}{2}}\widehat{Q}^{-\frac{1}{2}}\widehat{P}^{\frac{1}{2}}\widehat{Q}^{-\frac{1}{2}}\widehat{Q}^{\frac{1}{2}}\widehat{P}^{-\frac{1}{2}}\widehat{Q}^{\frac{1}{2}}\widehat{P}^{\frac{1}{2}}\widehat{Q}^{\frac{1}{2}}$.
For this assignment, we find the factor $\prod_{(i,j)}q^{\frac{\epsilon_i\epsilon_j}{4}}=q^{\frac{4-4}{4}}=1$ by taking the difference between the numbers of black and white boxes. 
}
\label{young}
\end{figure}

For the special case with only two types of branes, say $(1,k)$5-branes and NS5-branes, the two types correspond to the factors $\widehat{\cal P}=\widehat P^{\frac{1}{2}}+\widehat P^{-\frac{1}{2}}$ and $\widehat{\cal Q}=\widehat Q^{\frac{1}{2}}+\widehat Q^{-\frac{1}{2}}$.
In this case, we take $[\widehat q,\widehat p]=i\hbar$ with $\hbar=2\pi k$ ($k>0$).
In this setting, the computation can be carried out more systematically.
After fixing the total numbers of $(1,k)$5-branes $\widehat{\cal P}$ and NS5-branes $\widehat{\cal Q}$ to be $m$ and $n$, we associate each ordering of spectral operators with a Young diagram.
Namely, reading the operators from left to right, we move upward whenever we encounter a $(1,k)$5-brane factor $\widehat{\cal P}$, and move to the right whenever we encounter an NS5-brane factor $\widehat{\cal Q}$.
The resulting connected path determines the boundary of a Young diagram.
Strictly speaking, the ordering is encoded by a Young diagram inside an $m\times n$ rectangle.
Hence trailing zeros are retained, since they record the total number $m$ of $\widehat{\cal P}$ factors.
For example, when we have five $\widehat{\cal Q}$ factors and four $\widehat{\cal P}$ factors, the ordering $\widehat{\cal P}\widehat{\cal Q}\widehat{\cal P}\widehat{\cal Q}^2\widehat{\cal P}\widehat{\cal Q}\widehat{\cal P}\widehat{\cal Q}$ corresponds to the Young diagram $[4,3,1,0]$ (see figure \ref{young}).

Using the commutation relations
\begin{align}
\widehat Q\widehat P=q\widehat P\widehat Q,
\label{QPnc}
\end{align}
with $q=e^{i\hbar}$, we obtain the general expression
\begin{align}
\widehat H=\sum_{m',n'}q^{-\frac{m'n'}{8}}
\Bigg(\sum_{\{\pm\}}\prod_{(i,j)\in\lambda}q^{\frac{\epsilon_i\epsilon_j}{4}}\Bigg)
\big[\widehat P^{\frac{m'}{2}}\widehat Q^{\frac{n'}{2}}\big].
\end{align}
Namely, for the coefficients of $\big[\widehat P^{\frac{m'}{2}}\widehat Q^{\frac{n'}{2}}\big]$ coming from a spectral operator $\widehat H$ consisting of $m$ factors of $\widehat{\cal P}$ and $n$ factors of $\widehat{\cal Q}$, we need to choose $\frac{m+m'}{2}$ of $\widehat P^{\frac{1}{2}}$ and $\frac{m-m'}{2}$ of $\widehat P^{-\frac{1}{2}}$ out of $m$ factors of $\widehat{\cal P}=\widehat P^{\frac{1}{2}}+\widehat P^{-\frac{1}{2}}$ and choose $\frac{n+n'}{2}$ of $\widehat Q^{\frac{1}{2}}$ and $\frac{n-n'}{2}$ of $\widehat Q^{-\frac{1}{2}}$ out of $n$ factors of $\widehat{\cal Q}=\widehat Q^{\frac{1}{2}}+\widehat Q^{-\frac{1}{2}}$.
This corresponds to assigning a sign $\pm$ to each arrow along the boundary path (figure \ref{young}).
Then, by bringing operators $\widehat P^{\frac{\epsilon_i}{2}}$ to the left and $\widehat Q^{\frac{\epsilon_j}{2}}$ to the right using \eqref{QPnc}, we obtain factors $\prod_{(i,j)}q^{\frac{\epsilon_i\epsilon_j}{4}}$ depending on the product of the vertical $i$ and horizontal $j$ signs.
Finally, we restore the Weyl ordering $\widehat P^{\frac{m'}{2}}\widehat Q^{\frac{n'}{2}}=q^{-\frac{m'n'}{8}}\big[\widehat P^{\frac{m'}{2}}\widehat Q^{\frac{n'}{2}}\big]$ and sum over all the assignments of signs which contribute to the same powers of $\big[\widehat P^{\frac{m'}{2}}\widehat Q^{\frac{n'}{2}}\big]$.
 
As previously, coefficients associated with internal lattice points of the Newton polygon are generally nonlinear and the expression is complicated.
However, if we are only interested in the boundary points of the Newton polygon, we can express the relative coefficients (or the asymptotic values) in terms of the Young diagram data.
Using the commutation relations
\begin{align}
&\widehat Q^{\frac{\lambda}{2}}(\widehat P^{\frac{1}{2}}+\widehat P^{-\frac{1}{2}})
=(q^{\frac{\lambda}{4}}\widehat P^{\frac{1}{2}}+q^{-\frac{\lambda}{4}}\widehat P^{-\frac{1}{2}})\widehat Q^{\frac{\lambda}{2}},\quad
\widehat Q^{-\frac{\lambda}{2}}(\widehat P^{\frac{1}{2}}+\widehat P^{-\frac{1}{2}})
=(q^{-\frac{\lambda}{4}}\widehat P^{\frac{1}{2}}+q^{\frac{\lambda}{4}}\widehat P^{-\frac{1}{2}})\widehat Q^{-\frac{\lambda}{2}},\nonumber\\
&(\widehat Q^{\frac{1}{2}}+\widehat Q^{-\frac{1}{2}})\widehat P^{\frac{\lambda'}{2}}
=\widehat P^{\frac{\lambda'}{2}}(q^{\frac{\lambda'}{4}}\widehat Q^{\frac{1}{2}}+q^{-\frac{\lambda'}{4}}\widehat Q^{-\frac{1}{2}}),\quad
(\widehat Q^{\frac{1}{2}}+\widehat Q^{-\frac{1}{2}})\widehat P^{-\frac{\lambda'}{2}}
=\widehat P^{-\frac{\lambda'}{2}}(q^{-\frac{\lambda'}{4}}\widehat Q^{\frac{1}{2}}+q^{\frac{\lambda'}{4}}\widehat Q^{-\frac{1}{2}}),
\label{QPyoung}
\end{align}
we can bring all the $\widehat P$ operators to the left and all the $\widehat Q$ operators to the right.
After taking care of the factor from the Weyl ordering, we find the relative coefficients of the boundary terms are given by
\begin{eqnarray}
&\big\{q^{\frac{m}{4}-\frac{\lambda'_j}{2}}\big\}_{j=1}^n&\nonumber\\*
\big\{q^{-\frac{n}{4}+\frac{\lambda_i}{2}}\big\}_{i=1}^m\hspace{-6mm}&\raisebox{-4mm}{\scalebox{4}{$\Box$}}&\hspace{-6mm}\big\{q^{\frac{n}{4}-\frac{\lambda_i}{2}}\big\}_{i=1}^m\nonumber\\*
&\big\{q^{-\frac{m}{4}+\frac{\lambda'_j}{2}}\big\}_{j=1}^n&
\end{eqnarray}
for the quantum curve considered here.
Here we regard the right and left sides as the lines $Q=\infty$ and $Q=0$ respectively, while the upper and lower sides as $P=\infty$ and $P=0$. 
Then, the vertical and horizontal sides contain $m$ and $n$ asymptotic values respectively.

From the computations in this subsection, we have learned that the polytope arguments developed for brane configurations admit a natural reformulation in terms of the asymptotic values of quantum curves.
In the next sections, we shall turn to the study of polytopes with quantum curves, starting with the standard case of $\widehat D_5$ before turning to the more complicated cases of $\widehat E_6$ and $\widehat E_7$.

\section{$D_5$ curves}

In the previous section, we explained the fundamental domains of duality cascades in brane configurations and reformulated the finiteness and uniqueness problems as the question of whether these domains are parallelotopes.
In the absence of FI parameters, this was established through the zonotope description, and the vertices were identified with brane configurations of vanishing relative ranks.
We also observed that the same discussion admits a reformulation in terms of quantum curves, which are factorized precisely at the vertices.

Although the HW transitions are not always transparent for quantum curves associated with del Pezzo geometries with FI parameters, the exceptional Weyl group can still be used to generate all linear inequalities defining the fundamental domain \cite{FMMN}, yielding the higher-dimensional ${\cal H}$-description.
The lower-dimensional ${\cal V}$-description, however, has not yet been understood directly in terms of quantum curves.
Since the fundamental domain does not admit the powerful ${\cal Z}$-description, understanding the closely related ${\cal V}$-description becomes particularly important.

In this and the following section, we investigate the vertices of the fundamental domain from the viewpoint of quantum curves.
Interestingly, we find that the vertices correspond to completely factorized quantum curves constructed from $\widehat{\cal Q}^\pm$ and $\widehat{\cal P}^\pm$ \eqref{QPpm}.
More concretely, by applying the BCH formula \eqref{BCH} to quantum curves, we read off the asymptotic values of the curve and find that these asymptotic values coincide with the vertices obtained from the ${\cal H}$-description generated by the Weyl-group action.
We begin with the $D_5$ curve because its brane interpretation is the most transparent among the cases considered in this paper.

\subsection{Fundamental domain}

We first recapitulate the $D_5$ quantum curve, together with its associated brane configuration and effective gauge theory, and then turn to the fundamental domain in the ${\cal H}$-description.

\begin{figure}[!t]
\vspace{-12mm}
\centering\includegraphics[width=16cm]{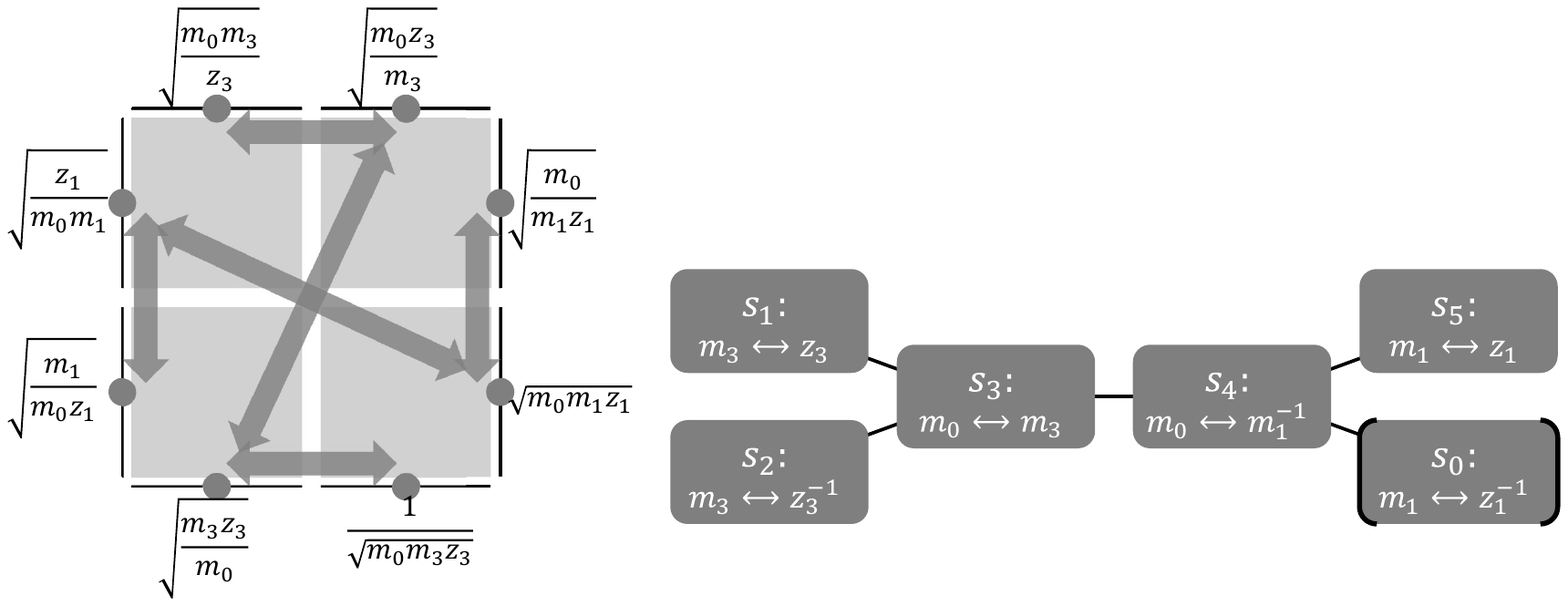}
\vspace{-36mm}
\caption{Asymptotic values and Weyl reflections for the $D_5$ quantum curve.
Asymptotic values \eqref{D5asymptotic} are schematically shown in the figure, and the Weyl reflections act by exchanging them.}
\label{d5fig}
\end{figure}

The $D_5$ quantum curve arises from a class of ${\cal N}=4$ supersymmetric Chern-Simons theories related to the ABJM theory (as explained shortly below).
In terms of the canonical operators $\widehat Q=e^{\widehat q}$ and $\widehat P=e^{\widehat p}$ satisfying $\widehat Q\widehat P=q\widehat P\widehat Q$ \eqref{QPnc} with $q=e^{i\hbar}$ and $\hbar=2\pi k$ $(k>0)$, it is given explicitly by
\begin{align}
&\widehat H/\alpha=q^{-\frac{1}{2}}\widehat Q(\widehat P+q^{\frac{1}{2}}\sqrt{m_0m_1z_1})\Big(\widehat P+q^{\frac{1}{2}}\sqrt{\frac{m_0}{m_1z_1}}\Big)\widehat 
P^{-1}\nonumber\\
&\quad+\Big(\sqrt{\frac{m_0z_3}{m_3}}+\sqrt{\frac{m_0m_3}{z_3}}\Big)\widehat P+E/\alpha+\Big(\sqrt{m_0m_3z_3}+\sqrt{\frac{m_0}{m_3z_3}}\Big)\widehat P^{-1}\nonumber\\
&\quad+q^{\frac{1}{2}}m_0\widehat Q^{-1}\Big(\widehat P+q^{-\frac{1}{2}}\sqrt{\frac{z_1}{m_0m_1}}\Big)\Big(\widehat P+q^{-\frac{1}{2}}\sqrt{\frac{m_1}{m_0z_1}}\Big)\widehat P^{-1},
\label{d5curve}
\end{align}
which is equivalently characterized by its asymptotic values \cite{MN9}
\begin{align}
&(\infty,-\sqrt{m_0m_1z_1}),\quad\Big(\infty,-\sqrt{\frac{m_0}{m_1z_1}}\Big),\quad\Big(-\sqrt{\frac{m_0z_3}{m_3}},\infty\Big),\quad\Big(-\sqrt{\frac{m_0m_3}{z_3}},\infty\Big),\nonumber\\
&\Big(0,-\sqrt{\frac{z_1}{m_0m_1}}\Big),\quad\Big(0,-\sqrt{\frac{m_1}{m_0z_1}}\Big),\quad\Big(-\sqrt{\frac{m_3z_3}{m_0}},0\Big),\quad\Big(-\frac{1}{\sqrt{m_0m_3z_3}},0\Big),
\label{D5asymptotic}
\end{align}
as schematically summarized in figure \ref{d5fig}.
We include the trivial $q$-dependence in \eqref{d5curve} so that the asymptotic values can be read off directly when the quantum curve is expressed in the Weyl ordering as in \eqref{E1234BCH}.
Note that, in discussing the asymptotic values of quantum curves, we do not distinguish small gauge transformations generated by the adjoint actions of $\widehat Q$ or $\widehat P$, which only rescale the vertical and horizontal asymptotic values.
For this reason, although there appear to be eight parameters in the asymptotic values, after removing two redundant overall factors together with one constraint from Vieta's formula, only five of them $(m_0,m_1,m_3,z_1,z_3)$ remain.
This curve is known to be invariant under the action of the $D_5$ Weyl group $W(D_5)$ (see figure \ref{d5fig}).
The Weyl reflections are realized either by trivial exchanges of asymptotic values or by similarity transformations of $\widehat H$ constructed in \cite{Has,Mor}.
The overall factor $\alpha$ is also fixed by $W(D_5)$, which we suppress to avoid unnecessary complications.

This $D_5$ curve consists of operator monomials $\widehat P^m\widehat Q^n$ ($m,n=\pm 1,0$), whose Newton polygon is a $2\times 2$ rectangle.
In the Fermi gas formalism, a $(1,k)$5-brane and an NS5-brane are represented by $\widehat{\cal P}$ and $\widehat{\cal Q}$ respectively, and each factor increases the extent of the Newton polygon in the corresponding direction by one unit.
Hence, the rectangular support of the $D_5$ curve naturally suggests a brane configuration with two $(1,k)$5-branes and two NS5-branes.
Indeed, if we cut the T-duality circle, denote the endpoints by brackets and represent $(1,k)$5-branes and NS5-branes by $\bullet$ and $\circ$ respectively, the corresponding brane configuration is given by \cite{MN9}
\begin{align}
{\langle}N\overunderset{1}{Z_1}{\bullet}N+M_0+M_1+k\overunderset{2}{0}{\bullet}N+2M_0+2k\overunderset{3}{Z_3}{\circ}N+M_0+M_3+k\overunderset{4}{0}{\circ}{\rangle},
\label{d5brane}
\end{align}
where $(M_0,M_1,M_3)$ denote the relative ranks and $(Z_1,Z_3)$ the FI parameters.
Here the multiplicative parameters $(m_0,m_1,m_3,z_1,z_3)$ in the quantum curve \eqref{d5curve} are related to the additive parameters in the brane configuration \eqref{d5brane} by exponentiation:
\begin{align}
m_0=e^{2\pi iM_0},\quad
m_1=e^{2\pi iM_1},\quad
m_3=e^{2\pi iM_3},\quad
z_1=e^{2\pi iZ_1},\quad
z_3=e^{2\pi iZ_3}.
\end{align}

The corresponding effective gauge theory has gauge group
\begin{align}
\text{U}(N)_{-k}\times\text{U}(N+M_0+M_1+k)_{0}\times\text{U}(N+2M_0+2k)_{k}\times\text{U}(N+M_0+M_3+k)_{0},
\end{align}
with the FI parameters $(Z_1,Z_3)$ included.
Here we shift relative ranks $(M_0,M_1,M_3)$ so that the origin is invariant under the action of $W(D_5)$.
The grand canonical partition function \eqref{GPF}, obtained by dualizing the overall rank $N$ in the partition function, is described, up to normalization, by the Fredholm determinant \eqref{Fredholm} of the $D_5$ quantum curve in \eqref{d5curve}.
Both the partition function and the grand canonical partition function have been studied extensively in \cite{MN3,MNN,MNY,KM,BGKNT,MN9,MNA}.

From the supersymmetry condition \cite{HW}, the numbers of D3-branes have to be nonnegative,
\begin{align}
-M_0-M_1\le k,\quad
-M_0-M_3\le k.
\end{align}
For the case of brane configurations without FI parameters, similar inequalities are obtained by requiring the numbers of D3-branes to remain nonnegative after applying the HW transitions arbitrarily.
For the current case with FI parameters, although the full set of inequalities are not immediately clear from the HW transitions, we can generate them with the $D_5$ Weyl group $W(D_5)$.
By acting with $W(D_5)$, we find $40$ inequalities
\begin{align}
\pm X_i\pm X_j\le k,
\label{D5facet}
\end{align}
where $X_i$ and $X_j$ are two distinct elements of $\{M_0,M_1,M_3,Z_1,Z_3\}$.
These inequalities define the facets of the fundamental domain in the ${\cal H}$-description.
Remarkably, one of the inequalities, $M_1+Z_1\le k$, coincides with the integrability condition for level-$k$ highest-weight representations of the affine algebra \cite{CFT},
\begin{align}
\bm\theta\cdot\bm\lambda\le k,
\label{integrable}
\end{align}
where $\bm\theta=-{\bm\alpha}_0$ is the highest root, as observed in \cite{FMMN}.
Here the simple roots for $D_5$, inducing the Weyl reflections shown in figure \ref{d5fig}, are
\begin{align}
\!\!\begin{array}{l}
{\bm\alpha}_1=(0,0,-1,0,1),\\
{\bm\alpha}_2=(0,0,-1,0,-1),
\end{array}
{\bm\alpha}_3=(-1,0,1,0,0),\;
{\bm\alpha}_4=(1,1,0,0,0),
\begin{array}{l}
{\bm\alpha}_5=(0,-1,0,1,0),\\
{\bm\alpha}_0=(0,-1,0,-1,0),
\end{array}
\end{align}
and all roots are given by vectors with two entries equal to $\pm 1$.
Since the brane interpretation is clear for the $D_5$ curve, many of these inequalities can also be reproduced directly from the HW transitions in the corresponding brane configuration \cite{FMMN}.
This provides a nontrivial consistency check of the Weyl-group construction of the fundamental domain.

The fundamental domain without FI parameters was shown to be a zonotope.
A zonotope is the Minkowski sum of line segments, which in the present case are generated by exchanges of 5-branes.
In contrast, the fundamental domain in the presence of FI parameters is no longer a zonotope.
Nevertheless, as we now show, it is precisely the Voronoi polytope associated with the $D_5$ root lattice. 
The Voronoi polytope of a lattice is defined as the set of points closer to the origin than to any other lattice point.
It is known to be a parallelotope, and hence tiles the entire parameter space by discrete translations.

Indeed, requiring that ${\bm\lambda}=(M_0,M_1,M_3,Z_1,Z_3)$ be at least as close to the origin as to any other root-lattice point $\bm\alpha$, namely ${\bm\lambda}^2\le(\bm\lambda-\bm\alpha)^2$ or
\begin{align}
\bm\alpha\cdot\bm\lambda\le\bm\alpha^2/2,
\label{voronoi}
\end{align}
we obtain the inequalities $\pm X_i\pm X_j\le 1$, which reproduce precisely \eqref{D5facet} after rescaling by $k$.

\subsection{Vertices}

From the inequalities \eqref{D5facet}, it is tedious but straightforward to obtain the vertices
\begin{align}
&k(\pm 1/2,\pm 1/2,\pm 1/2,\pm 1/2,\pm 1/2),\label{D5vertices}\\
&k(\pm 1,0,0,0,0),\quad k(0,\pm 1,0,0,0),\quad k(0,0,\pm 1,0,0),\quad k(0,0,0,\pm 1,0),\quad k(0,0,0,0,\pm 1),\nonumber
\end{align}
in the ${\cal V}$-description.
Since the facets in \eqref{D5facet} are generated by the $D_5$ Weyl group, the resulting vertices are naturally organized into Weyl orbits of $D_5$.
Indeed, it is easy to see that the first line of \eqref{D5vertices} consists of the weights of the spinor and cospinor representations $\bm{16}$ and $\overline{\bm{16}}$, while the second line consists of the weights of the vector representation $\bm{10}$.
These are precisely the minuscule representations of $D_5$ \cite{Hum}, for which the Weyl group acts transitively on the weights.
This is consistent with the distinguished role often played by minuscule weights in the geometry of Voronoi polytopes of simply-laced root lattices \cite{CS}.

\begin{table}[t!]
\begin{center}
\begin{tabular}{c|c||c|c}
quantum curves&$(M_0,M_1,M_3,Z_1,Z_3)/k$&quantum curves&$(M_0,M_1,M_3,Z_1,Z_3)/k$\\\hline\hline
$\widehat{\cal Q}^-_4\widehat{\cal P}^-_2\widehat{\cal Q}^+_3\widehat{\cal P}^+_1$&
$(-\frac{1}{2},-\frac{1}{2},-\frac{1}{2},-\frac{1}{2},-\frac{1}{2})$&
$\widehat{\cal P}^+_1\widehat{\cal Q}^+_3\widehat{\cal P}^-_2\widehat{\cal Q}^-_4$&
$(+\frac{1}{2},+\frac{1}{2},+\frac{1}{2},+\frac{1}{2},+\frac{1}{2})$
\\
$\widehat{\cal Q}^-_4\widehat{\cal P}^+_2\widehat{\cal Q}^+_3\widehat{\cal P}^-_1$&
$(-\frac{1}{2},-\frac{1}{2},-\frac{1}{2},-\frac{1}{2},+\frac{1}{2})$&
$\widehat{\cal P}^-_1\widehat{\cal Q}^+_3\widehat{\cal P}^+_2\widehat{\cal Q}^-_4$&
$(+\frac{1}{2},+\frac{1}{2},+\frac{1}{2},+\frac{1}{2},-\frac{1}{2})$
\\
$\widehat{\cal Q}^+_4\widehat{\cal P}^-_2\widehat{\cal Q}^-_3\widehat{\cal P}^+_1$&
$(-\frac{1}{2},-\frac{1}{2},-\frac{1}{2},+\frac{1}{2},-\frac{1}{2})$&
$\widehat{\cal P}^+_1\widehat{\cal Q}^-_3\widehat{\cal P}^-_2\widehat{\cal Q}^+_4$&
$(+\frac{1}{2},+\frac{1}{2},+\frac{1}{2},-\frac{1}{2},+\frac{1}{2})$
\\
$\widehat{\cal Q}^+_4\widehat{\cal P}^+_2\widehat{\cal Q}^-_3\widehat{\cal P}^-_1$&
$(-\frac{1}{2},-\frac{1}{2},-\frac{1}{2},+\frac{1}{2},+\frac{1}{2})$&
$\widehat{\cal P}^-_1\widehat{\cal Q}^-_3\widehat{\cal P}^+_2\widehat{\cal Q}^+_4$&
$(+\frac{1}{2},+\frac{1}{2},+\frac{1}{2},-\frac{1}{2},-\frac{1}{2})$
\\\hline
$\widehat{\cal Q}^-_3\widehat{\cal P}^+_2\widehat{\cal Q}^+_4\widehat{\cal P}^-_1$&
$(-\frac{1}{2},-\frac{1}{2},+\frac{1}{2},-\frac{1}{2},-\frac{1}{2})$&
$\widehat{\cal P}^-_1\widehat{\cal Q}^+_4\widehat{\cal P}^+_2\widehat{\cal Q}^-_3$&
$(+\frac{1}{2},+\frac{1}{2},-\frac{1}{2},+\frac{1}{2},+\frac{1}{2})$
\\
$\widehat{\cal Q}^-_3\widehat{\cal P}^-_2\widehat{\cal Q}^+_4\widehat{\cal P}^+_1$&
$(-\frac{1}{2},-\frac{1}{2},+\frac{1}{2},-\frac{1}{2},+\frac{1}{2})$&
$\widehat{\cal P}^+_1\widehat{\cal Q}^+_4\widehat{\cal P}^-_2\widehat{\cal Q}^-_3$&
$(+\frac{1}{2},+\frac{1}{2},-\frac{1}{2},+\frac{1}{2},-\frac{1}{2})$
\\
$\widehat{\cal Q}^+_3\widehat{\cal P}^+_2\widehat{\cal Q}^-_4\widehat{\cal P}^-_1$&
$(-\frac{1}{2},-\frac{1}{2},+\frac{1}{2},+\frac{1}{2},-\frac{1}{2})$&
$\widehat{\cal P}^-_1\widehat{\cal Q}^-_4\widehat{\cal P}^+_2\widehat{\cal Q}^+_3$&
$(+\frac{1}{2},+\frac{1}{2},-\frac{1}{2},-\frac{1}{2},+\frac{1}{2})$
\\
$\widehat{\cal Q}^+_3\widehat{\cal P}^-_2\widehat{\cal Q}^-_4\widehat{\cal P}^+_1$&
$(-\frac{1}{2},-\frac{1}{2},+\frac{1}{2},+\frac{1}{2},+\frac{1}{2})$&
$\widehat{\cal P}^+_1\widehat{\cal Q}^-_4\widehat{\cal P}^-_2\widehat{\cal Q}^+_3$&
$(+\frac{1}{2},+\frac{1}{2},-\frac{1}{2},-\frac{1}{2},-\frac{1}{2})$
\\\hline
$\widehat{\cal Q}^+_4\widehat{\cal P}^-_1\widehat{\cal Q}^-_3\widehat{\cal P}^+_2$&
$(-\frac{1}{2},+\frac{1}{2},-\frac{1}{2},-\frac{1}{2},-\frac{1}{2})$&
$\widehat{\cal P}^+_2\widehat{\cal Q}^-_3\widehat{\cal P}^-_1\widehat{\cal Q}^+_4$&
$(+\frac{1}{2},-\frac{1}{2},+\frac{1}{2},+\frac{1}{2},+\frac{1}{2})$
\\
$\widehat{\cal Q}^+_4\widehat{\cal P}^+_1\widehat{\cal Q}^-_3\widehat{\cal P}^-_2$&
$(-\frac{1}{2},+\frac{1}{2},-\frac{1}{2},-\frac{1}{2},+\frac{1}{2})$&
$\widehat{\cal P}^-_2\widehat{\cal Q}^-_3\widehat{\cal P}^+_1\widehat{\cal Q}^+_4$&
$(+\frac{1}{2},-\frac{1}{2},+\frac{1}{2},+\frac{1}{2},-\frac{1}{2})$
\\
$\widehat{\cal Q}^-_4\widehat{\cal P}^-_1\widehat{\cal Q}^+_3\widehat{\cal P}^+_2$&
$(-\frac{1}{2},+\frac{1}{2},-\frac{1}{2},+\frac{1}{2},-\frac{1}{2})$&
$\widehat{\cal P}^+_2\widehat{\cal Q}^+_3\widehat{\cal P}^-_1\widehat{\cal Q}^-_4$&
$(+\frac{1}{2},-\frac{1}{2},+\frac{1}{2},-\frac{1}{2},+\frac{1}{2})$
\\
$\widehat{\cal Q}^-_4\widehat{\cal P}^+_1\widehat{\cal Q}^+_3\widehat{\cal P}^-_2$&
$(-\frac{1}{2},+\frac{1}{2},-\frac{1}{2},+\frac{1}{2},+\frac{1}{2})$&
$\widehat{\cal P}^-_2\widehat{\cal Q}^+_3\widehat{\cal P}^+_1\widehat{\cal Q}^-_4$&
$(+\frac{1}{2},-\frac{1}{2},+\frac{1}{2},-\frac{1}{2},-\frac{1}{2})$
\\\hline
$\widehat{\cal Q}^+_3\widehat{\cal P}^+_1\widehat{\cal Q}^-_4\widehat{\cal P}^-_2$&
$(-\frac{1}{2},+\frac{1}{2},+\frac{1}{2},-\frac{1}{2},-\frac{1}{2})$&
$\widehat{\cal P}^-_2\widehat{\cal Q}^-_4\widehat{\cal P}^+_1\widehat{\cal Q}^+_3$&
$(+\frac{1}{2},-\frac{1}{2},-\frac{1}{2},+\frac{1}{2},+\frac{1}{2})$
\\
$\widehat{\cal Q}^+_3\widehat{\cal P}^-_1\widehat{\cal Q}^-_4\widehat{\cal P}^+_2$&
$(-\frac{1}{2},+\frac{1}{2},+\frac{1}{2},-\frac{1}{2},+\frac{1}{2})$&
$\widehat{\cal P}^+_2\widehat{\cal Q}^-_4\widehat{\cal P}^-_1\widehat{\cal Q}^+_3$&
$(+\frac{1}{2},-\frac{1}{2},-\frac{1}{2},+\frac{1}{2},-\frac{1}{2})$
\\
$\widehat{\cal Q}^-_3\widehat{\cal P}^+_1\widehat{\cal Q}^+_4\widehat{\cal P}^-_2$&
$(-\frac{1}{2},+\frac{1}{2},+\frac{1}{2},+\frac{1}{2},-\frac{1}{2})$&
$\widehat{\cal P}^-_2\widehat{\cal Q}^+_4\widehat{\cal P}^+_1\widehat{\cal Q}^-_3$&
$(+\frac{1}{2},-\frac{1}{2},-\frac{1}{2},-\frac{1}{2},+\frac{1}{2})$
\\
$\widehat{\cal Q}^-_3\widehat{\cal P}^-_1\widehat{\cal Q}^+_4\widehat{\cal P}^+_2$&
$(-\frac{1}{2},+\frac{1}{2},+\frac{1}{2},+\frac{1}{2},+\frac{1}{2})$&
$\widehat{\cal P}^+_2\widehat{\cal Q}^+_4\widehat{\cal P}^-_1\widehat{\cal Q}^-_3$&
$(+\frac{1}{2},-\frac{1}{2},-\frac{1}{2},-\frac{1}{2},-\frac{1}{2})$
\\\hline
$\widehat{\cal Q}^\pm_{34}\widehat{\cal Q}^\mp_{43}\widehat{\cal P}^\pm_{12}\widehat{\cal P}^\mp_{21}$&
$(-1,0,0,0,0)$&
$\widehat{\cal P}^\pm_{12}\widehat{\cal P}^\mp_{21}\widehat{\cal Q}^\pm_{34}\widehat{\cal Q}^\mp_{43}$&
$(+1,0,0,0,0)$
\\
$\widehat{\cal P}^\pm_{2}\widehat{\cal Q}^\pm_{34}\widehat{\cal Q}^\mp_{43}\widehat{\cal P}^\mp_{1}$&
$(0,-1,0,0,0)$&
$\widehat{\cal P}^\pm_{1}\widehat{\cal Q}^\pm_{34}\widehat{\cal Q}^\mp_{43}\widehat{\cal P}^\mp_{2}$&
$(0,+1,0,0,0)$
\\
$\widehat{\cal Q}^\pm_{4}\widehat{\cal P}^\pm_{12}\widehat{\cal P}^\mp_{21}\widehat{\cal Q}^\mp_{3}$&
$(0,0,-1,0,0)$&
$\widehat{\cal Q}^\pm_{3}\widehat{\cal P}^\pm_{12}\widehat{\cal P}^\mp_{21}\widehat{\cal Q}^\mp_{4}$&
$(0,0,+1,0,0)$
\end{tabular}
\end{center}
\caption{Parameters $(M_0,M_1,M_3,Z_1,Z_3)$ identified from the factorized quantum curves.
The factorized curves listed in the first $16$ rows, built from alternating canonical operators, $\widehat{\cal Q}^\pm\widehat{\cal P}^\pm\widehat{\cal Q}^\mp\widehat{\cal P}^\mp$ and $\widehat{\cal P}^\pm\widehat{\cal Q}^\pm\widehat{\cal P}^\mp\widehat{\cal Q}^\mp$, reproduce the vertices belonging to the spinor ${\bf 16}$ and cospinor $\overline{\bf 16}$ representations in \eqref{D5vertices}.
The curves in the last three rows, containing adjacent pairs of canonical operators such as $\widehat{\cal Q}^\pm\widehat{\cal Q}^\mp\widehat{\cal P}^\pm\widehat{\cal P}^\mp$, $\widehat{\cal P}^\pm\widehat{\cal P}^\mp\widehat{\cal Q}^\pm\widehat{\cal Q}^\mp$ and $\widehat{\cal P}^\pm\widehat{\cal Q}^\pm\widehat{\cal Q}^\mp\widehat{\cal P}^\mp$, reproduce the vertices in the vector representation ${\bf 10}$.
Note that the vertices in the vector representation admit multiple realizations, since $\widehat{\cal O}^+$ and $\widehat{\cal O}^-$ commute and $\widehat{\cal O}^+\cdots\widehat{\cal O}^-$ and $\widehat{\cal O}^-\cdots\widehat{\cal O}^+$ are related by small gauge transformations.
We therefore list multiple equivalent realizations whenever possible.
}
\label{D5vertex1}
\end{table}

In the case without FI parameters, the study of the finiteness and uniqueness of duality cascades \cite{FMS} relied crucially on establishing the equivalence of the three descriptions, ${\cal V}={\cal Z}={\cal H}$, and on verifying the space-filling property of the corresponding zonotope.
One of the key facts in the analysis is that the vertices in the ${\cal V}$-description are identified as brane configurations with vanishing relative ranks but in various orders of 5-branes.
In terms of quantum curves, we have seen that the condition of vanishing relative ranks is translated into factorized quantum curves.

Motivated by this observation, we now seek an interpretation of the vertices in the presence of FI parameters directly in terms of quantum curves.
We find that the vertices are precisely characterized by factorized quantum curves built from $\widehat{\cal P}^\pm$ and $\widehat{\cal Q}^\pm$ \eqref{QPpm}.
We shall spell out the details in this section.

We begin by considering various orderings of these canonical factors.
For the quantum curve to fall into the $D_5$ quantum curve, we need to choose a pair of $\widehat{\cal P}^\pm$ and a pair of $\widehat{\cal Q}^\pm$, corresponding to two $(1,k)$5-branes and two NS5-branes.
We label $\widehat{\cal P}^\pm$ by $1,2$ and $\widehat{\cal Q}^\pm$ by $3,4$, where $\widehat{\cal P}^\pm_1$ and $\widehat{\cal P}^\pm_2$ correspond to the asymptotic values $\sqrt{m_0m_1z_1}$, $\sqrt{z_1/(m_0m_1)}$ and $\sqrt{m_0/(m_1z_1)}$, $\sqrt{m_1/(m_0z_1)}$ in \eqref{D5asymptotic} respectively, while $\widehat{\cal Q}^\pm_3$ and $\widehat{\cal Q}^\pm_4$ correspond to $\sqrt{m_0z_3/m_3}$, $\sqrt{m_3z_3/m_0}$ and $\sqrt{m_0m_3/z_3}$, $1/\sqrt{m_0m_3z_3}$ respectively.
In other words, the labels serve to distinguish asymptotic values on the same side, since each asymptotic value can be associated with the canonical operator from which it originates through \eqref{QPyoung}.
For example, let us consider $\widehat{\cal Q}^-_4\widehat{\cal P}^-_2\widehat{\cal Q}^+_3\widehat{\cal P}^+_1$.
Applying the reduced BCH formula \eqref{QPyoung}, dressed with $\widehat Q^{\pm\frac{1}{2}}$ and $\widehat P^{\pm\frac{1}{2}}$, we obtain the asymptotic values $(\infty,-q^{-\frac{1}{2}})$, $(\infty,-q^{\frac{1}{2}})$, $(-q^{-\frac{1}{2}},\infty)$, $(-q^{-\frac{1}{2}},\infty)$, $(0,-q^{\frac{1}{2}})$, $(0,-q^{\frac{1}{2}})$, $(-q^{-\frac{1}{2}},0)$, $(-q^{\frac{1}{2}},0)$.
Comparing these asymptotic values with \eqref{D5asymptotic}, we can read off the corresponding parameters $(M_0,M_1,M_3,Z_1,Z_3)$.
The results are summarized in table \ref{D5vertex1}.
It turns out that the vertices \eqref{D5vertices} in the spinor ${\bf 16}$ and cospinor $\overline{\bf 16}$ representations correspond precisely to factorized curves in which $\widehat{\cal Q}^\pm$ and $\widehat{\cal P}^\pm$ appear alternately, while the vertices in the vector representation ${\bf 10}$ arise when identical canonical factors appear adjacently in pairs.
While the alternating configurations lead to distinct vertices, different orderings containing adjacent identical factors may yield the same vertex, resulting in the degeneracies observed in the ${\bf 10}$ representation (table \ref{D5vertex1}).

\begin{table}[t!]
\begin{center}
\begin{tabular}{c|c||c|c}
separations&facet&separations&facet\\\hline\hline
$\{\widehat{\cal P}_2,\widehat{\cal Q}_3,\widehat{\cal Q}_4\}\cdot\{\widehat{\cal P}_1\}$&$-M_0-M_1\le k$&
$\{\widehat{\cal P}_1\}\cdot\{\widehat{\cal P}_2,\widehat{\cal Q}_3,\widehat{\cal Q}_4\}$&$M_0+M_1\le k$\\
$\{\widehat{\cal P}_1,\widehat{\cal Q}_3,\widehat{\cal Q}_4\}\cdot\{\widehat{\cal P}_2\}$&$-M_0+M_1\le k$&
$\{\widehat{\cal P}_2\}\cdot\{\widehat{\cal P}_1,\widehat{\cal Q}_3,\widehat{\cal Q}_4\}$&$M_0-M_1\le k$\\
$\{\widehat{\cal Q}_4\}\cdot\{\widehat{\cal P}_1,\widehat{\cal P}_2,\widehat{\cal Q}_3\}$&$-M_0-M_3\le k$&
$\{\widehat{\cal P}_1,\widehat{\cal P}_2,\widehat{\cal Q}_3\}\cdot\{\widehat{\cal Q}_4\}$&$M_0+M_3\le k$\\
$\{\widehat{\cal Q}_3\}\cdot\{\widehat{\cal P}_1,\widehat{\cal P}_2,\widehat{\cal Q}_4\}$&$-M_0+M_3\le k$&
$\{\widehat{\cal P}_1,\widehat{\cal P}_2,\widehat{\cal Q}_4\}\cdot\{\widehat{\cal Q}_3\}$&$M_0-M_3\le k$\\
$\{\widehat{\cal P}_2,\widehat{\cal Q}_4\}\cdot\{\widehat{\cal P}_1,\widehat{\cal Q}_3\}$&$-M_1-M_3\le k$&
$\{\widehat{\cal P}_1,\widehat{\cal Q}_3\}\cdot\{\widehat{\cal P}_2,\widehat{\cal Q}_4\}$&$M_1+M_3\le k$\\
$\{\widehat{\cal P}_2,\widehat{\cal Q}_3\}\cdot\{\widehat{\cal P}_1,\widehat{\cal Q}_4\}$&$-M_1+M_3\le k$&
$\{\widehat{\cal P}_1,\widehat{\cal Q}_4\}\cdot\{\widehat{\cal P}_2,\widehat{\cal Q}_3\}$&$M_1-M_3\le k$
\end{tabular}
\end{center}
\caption{Correspondence between facets and separations of factorized curves.
When the inequality defining a facet involves only relative ranks $(M_0,M_1,M_3)$, the interpretation from the case without FI parameters remains valid.
}
\label{D5facet1}
\end{table}

Interestingly, almost the entire set of vertices in \eqref{D5vertices} is reproduced simply by considering various orderings of $\widehat{\cal P}^\pm$ and $\widehat{\cal Q}^\pm$ listed in table \ref{D5vertex1}.
This provides strong evidence for identifying the vertices in \eqref{D5vertices} with factorized quantum curves.
Apparently, four vertices remain unidentified:
\begin{align}
(M_0,M_1,M_3,Z_1,Z_3)=(0,0,0,\pm k,0),\;(0,0,0,0,\pm k).
\label{D5unid}
\end{align}
which involve only asymptotic values associated with the FI parameters $(Z_1,Z_3)$.
One possible interpretation is that these correspond to a coarse-grained limit in which the 5-branes are effectively smeared out, leaving only nontrivial FI parameters.
Note, however, that this interpretation should not be taken too literally.
Even for these vertices, well-defined brane configurations can be constructed without difficulty.
We only mean that the configuration with vanishing $(M_0,M_1,M_3)$ may naturally be regarded as the average over configurations with all possible orderings.
We stress again that, except for these unidentified vertices, the asymptotic values obtained from the factorized quantum curves coincide with the vertices associated with the minuscule representations $\bm{16}$, $\overline{\bm{16}}$ and $\bm{10}$.

To support this claim, we have listed all corresponding factorized quantum curves in table \ref{D5vertex1}.
The reader may wonder why this is necessary, given that all vertices are completely determined by the Weyl group $W(D_5)$.
It is, of course, straightforward to generate the full Weyl orbit from a single weight in each representation.
However, the Weyl-group action on the factorized quantum curves is not very clear.
We hope that our table is also helpful for understanding the Weyl-group action on the quantum curves.

For the case without FI parameters, duality cascades changing the reference interval by the cyclic symmetry are realized as discrete translations from one facet to the opposite one \cite{FMS}.
This remains true for inequalities involving only the relative ranks $(M_0,M_1,M_3)$.
Namely, the vertices corresponding to factorizations that separate the 5-branes into two identical sets lie on the same facet.
For example, all the configurations in which $\widehat{\cal P}_1$ is separated from the remaining factors, namely $\widehat{\cal Q}_4\widehat{\cal P}_2\widehat{\cal Q}_3\widehat{\cal P}_1$, $\widehat{\cal Q}_3\widehat{\cal P}_2\widehat{\cal Q}_4\widehat{\cal P}_1$, $\widehat{\cal Q}_3\widehat{\cal Q}_4\widehat{\cal P}_1\widehat{\cal P}_2$ and $\widehat{\cal P}_2\widehat{\cal Q}_3\widehat{\cal Q}_4\widehat{\cal P}_1$ reside on the facet $-M_0-M_1\le k$ (see table \ref{D5facet1}).
Then, moving the separated factor $\widehat{\cal P}_1$ from the rightmost to the leftmost consistently corresponds to the discrete translation $k(1,1,0,0,0)$.

\begin{table}[t!]
\begin{center}
\begin{tabular}{c|c||c|c}
patterns&facets&patterns&facets\\\hline\hline
$\widehat{\cal Q}^+\widehat{\cal P}_1\widehat{\cal Q}^-\widehat{\cal P}_2$, $\widehat{\cal Q}^-\widehat{\cal P}_2\widehat{\cal Q}^+\widehat{\cal P}_1$&$-M_0-Z_1\le k$&
$\widehat{\cal P}_1\widehat{\cal Q}^+\widehat{\cal P}_2\widehat{\cal Q}^-$, $\widehat{\cal P}_2\widehat{\cal Q}^-\widehat{\cal P}_1\widehat{\cal Q}^+$&$M_0+Z_1\le k$\\
$\widehat{\cal Q}^-\widehat{\cal P}_1\widehat{\cal Q}^+\widehat{\cal P}_2$, $\widehat{\cal Q}^+\widehat{\cal P}_2\widehat{\cal Q}^-\widehat{\cal P}_1$&$-M_0+Z_1\le k$&
$\widehat{\cal P}_1\widehat{\cal Q}^-\widehat{\cal P}_2\widehat{\cal Q}^+$, $\widehat{\cal P}_2\widehat{\cal Q}^+\widehat{\cal P}_1\widehat{\cal Q}^-$&$M_0-Z_1\le k$\\
$\widehat{\cal Q}_3\widehat{\cal P}^+\widehat{\cal Q}_4\widehat{\cal P}^-$, $\widehat{\cal Q}_4\widehat{\cal P}^-\widehat{\cal Q}_3\widehat{\cal P}^+$&$-M_0-Z_3\le k$&
$\widehat{\cal P}^+\widehat{\cal Q}_3\widehat{\cal P}^-\widehat{\cal Q}_4$, $\widehat{\cal P}^-\widehat{\cal Q}_4\widehat{\cal P}^+\widehat{\cal Q}_3$&$M_0+Z_3\le k$\\
$\widehat{\cal Q}_3\widehat{\cal P}^-\widehat{\cal Q}_4\widehat{\cal P}^+$, $\widehat{\cal Q}_4\widehat{\cal P}^+\widehat{\cal Q}_3\widehat{\cal P}^-$&$-M_0+Z_3\le k$&
$\widehat{\cal P}^-\widehat{\cal Q}_3\widehat{\cal P}^+\widehat{\cal Q}_4$, $\widehat{\cal P}^+\widehat{\cal Q}_4\widehat{\cal P}^-\widehat{\cal Q}_3$&$M_0-Z_3\le k$\\
$\widehat{\cal P}_2\widehat{\cal Q}^+\widehat{\cal P}_1$&
$-M_1-Z_1\le k$&
$\widehat{\cal P}_1\widehat{\cal Q}^+\widehat{\cal P}_2$&
$M_1+Z_1\le k$\\
$\widehat{\cal P}_2\widehat{\cal Q}^-\widehat{\cal P}_1$&
$-M_1+Z_1\le k$&
$\widehat{\cal P}_1\widehat{\cal Q}^-\widehat{\cal P}_2$&
$M_1-Z_1\le k$\\
$\widehat{\cal P}^-_2\widehat{\cal Q}_3\widehat{\cal P}^+_1$,
$\widehat{\cal P}^+_2\widehat{\cal Q}_4\widehat{\cal P}^-_1$&
$-M_1-Z_3\le k$&
$\widehat{\cal P}^+_1\widehat{\cal Q}_3\widehat{\cal P}^-_2$,
$\widehat{\cal P}^-_1\widehat{\cal Q}_4\widehat{\cal P}^+_2$&
$M_1+Z_3\le k$\\
$\widehat{\cal P}^-_2\widehat{\cal Q}_4\widehat{\cal P}^+_1$,
$\widehat{\cal P}^+_2\widehat{\cal Q}_3\widehat{\cal P}^-_1$&
$-M_1+Z_3\le k$&
$\widehat{\cal P}^+_1\widehat{\cal Q}_4\widehat{\cal P}^-_2$,
$\widehat{\cal P}^-_1\widehat{\cal Q}_3\widehat{\cal P}^+_2$&
$M_1-Z_3\le k$\\
$\widehat{\cal Q}^-_4\widehat{\cal P}_2\widehat{\cal Q}^+_3$,
$\widehat{\cal Q}^+_4\widehat{\cal P}_1\widehat{\cal Q}^-_3$&
$-M_3-Z_1\le k$&
$\widehat{\cal Q}^-_3\widehat{\cal P}_1\widehat{\cal Q}^+_4$,
$\widehat{\cal Q}^+_3\widehat{\cal P}_2\widehat{\cal Q}^-_4$&
$M_3+Z_1\le k$\\
$\widehat{\cal Q}^-_4\widehat{\cal P}_1\widehat{\cal Q}^+_3$,
$\widehat{\cal Q}^+_4\widehat{\cal P}_2\widehat{\cal Q}^-_3$&
$-M_3+Z_1\le k$&
$\widehat{\cal Q}^-_3\widehat{\cal P}_2\widehat{\cal Q}^+_4$,
$\widehat{\cal Q}^+_3\widehat{\cal P}_1\widehat{\cal Q}^-_4$&
$M_3-Z_1\le k$\\
$\widehat{\cal Q}_4\widehat{\cal P}^-\widehat{\cal Q}_3$&
$-M_3-Z_3\le k$&
$\widehat{\cal Q}_3\widehat{\cal P}^-\widehat{\cal Q}_4$&
$M_3+Z_3\le k$\\
$\widehat{\cal Q}_4\widehat{\cal P}^+\widehat{\cal Q}_3$&
$-M_3+Z_3\le k$&
$\widehat{\cal Q}_3\widehat{\cal P}^+\widehat{\cal Q}_4$&
$M_3-Z_3\le k$\\
\end{tabular}
\end{center}
\caption{Patterns of factorized curves on facets involving the FI parameters $(Z_1,Z_3)$, corresponding to the ${\bf 16}$ and $\overline{\bf 16}$ representations.}
\label{D5facet2}
\end{table}

For the inequalities involving the FI parameters $(Z_1,Z_3)$, however, the interpretation of the discrete translations between opposite facets in terms of duality cascades is less clear.
We only observe that, in the presence of FI parameters, the factorized curves lying on a given facet and belonging to the spinor representations exhibit similar patterns, as listed in table \ref{D5facet2}.
The meaning of these discrete translations, however, remains to be clarified.

\section{Curves with asymptotic degeneracies}

After studying the $D_5$ curve, let us turn to the $E_6$ and $E_7$ cases with degeneracies.
As emphasized in the introduction, the gauge-theoretic interpretations of these curves are not yet fully understood.
Nevertheless, their partition functions exhibit the Airy-function behavior characteristic of M2-branes.
Moreover, their exceptional Weyl-group symmetries suggest that they should satisfy the corresponding $q$-Painlev\'e equations, as in the ABJM theory and the $D_5$ case \cite{BGT,BGKNT,MN9,MNA}.
These observations motivate a closer investigation of these curves.
We shall show that, despite the absence of a transparent brane interpretation, the minuscule vertices of the fundamental domain can still be realized by factorized quantum curves constructed from $\widehat{\cal Q}^\pm$ and $\widehat{\cal P}^\pm$.
This provides a remarkable characterization of the minuscule vertices in terms of factorized quantum curves.

\subsection{$E_6$}

\begin{figure}[!t]
\vspace{-12mm}
\centering\includegraphics[width=16cm]{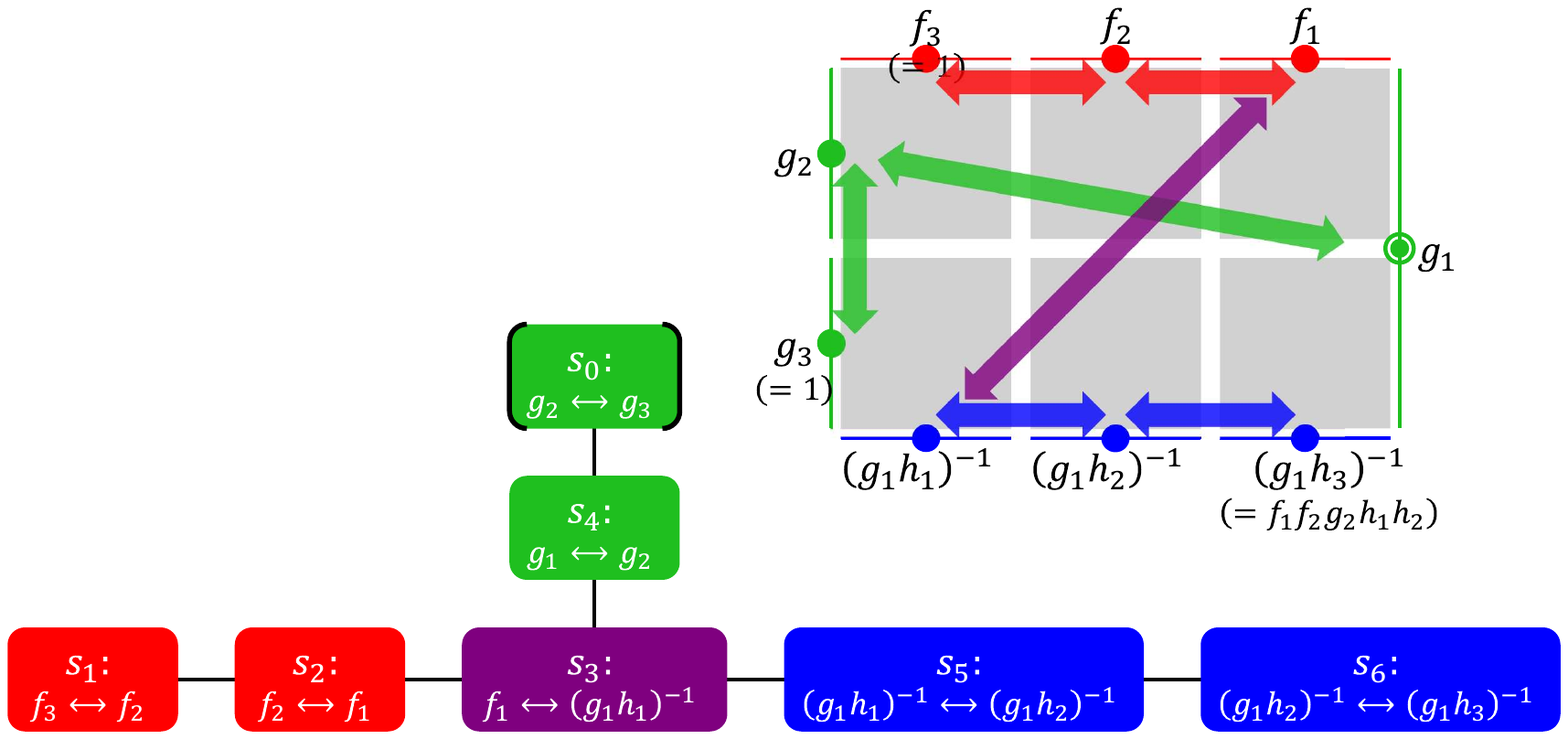}
\vspace{-24mm}
\caption{Asymptotic values and Weyl reflections for the $E_6$ quantum curve.}
\label{e6fig}
\end{figure}

For these exceptional curves we follow the convention of \cite{Mor}.
There the $E_6$ quantum curve, in the rectangular realization with support $\widehat P^m\widehat Q^n$ ($-1\le m\le 1,-1\le n\le 2$) is given by
\begin{align}
&\widehat H/\alpha=q^{-1}\widehat Q^2(\widehat P+q^{\frac{3}{2}}g_1)(\widehat P+q^{\frac{1}{2}}g_1)\widehat P^{-1}\nonumber\\
&\quad+q^{-\frac{1}{2}}\widehat Q(\widehat P+q^{\frac{1}{2}}g_1)((f_1+f_2+f_3)\widehat P+q^{\frac{1}{2}}(h_1^{-1}+h_2^{-1}+h_3^{-1}))\widehat P^{-1}\nonumber\\
&\quad+f_1f_2f_3(f_1^{-1}+f_2^{-1}+f_3^{-1})\widehat P+E/\alpha+(h_1h_2h_3)^{-1}(h_1+h_2+h_3)\widehat P^{-1}\nonumber\\
&\quad+q^{\frac{1}{2}}f_1f_2f_3\widehat Q^{-1}(\widehat P+q^{-\frac{1}{2}}g_2)(\widehat P+q^{-\frac{1}{2}}g_3)\widehat P^{-1},
\label{e6curve}
\end{align}
with the parameters $f_i$, $g_i$ and $h_i$ satisfying $f_1f_2f_3g_1g_2g_3h_1h_2h_3=1$.
The asymptotic values and the Weyl reflections for this curve are given in figure \ref{e6fig} using the shorthand notation introduced previously.

Classically, the $E_6$ curve is known \cite{BBT,KY} to be the most general toric curve, realized as a cubic Laurent polynomial on a triangular Newton polygon without degeneracies.
When the curve is realized on a rectangular Newton polygon by a symplectic transformation, one of the asymptotic values, $g_1$, becomes doubly degenerate.
For quantum curves, in \cite{Tak,Mor,MY} it was found that the classical degeneracy is split into asymptotic values $-q^{\frac{3}{2}}g_1$ and $-q^{\frac{1}{2}}g_1$ (or, $-q^{\frac{1}{2}}g_1$ and $-q^{-\frac{1}{2}}g_1$ in the Weyl ordering) differing by a factor of $q$ as can be read off from the $\widehat Q^2$ terms in \eqref{e6curve}.

Since we do not distinguish quantum curves related by small gauge transformations, we often fix the gauge ambiguity in the parameter space by setting the last parameters among $f_i$ and $g_i$ to unity, with the last parameter of $h_i$ then determined by the constraint as \cite{Mor}
\begin{align}
f_3=g_3=1,\quad h_3=(f_1f_2g_1g_2h_1h_2)^{-1}.
\end{align}
Note that, after gauge fixing, one must generally use small gauge transformations to restore the gauge-fixed form after applying Weyl reflections.
From these Weyl reflections, we can read off the simple roots and the fundamental weights of $E_6$, which are summarized in appendix \ref{rootweight} for completeness.
We need to be careful, however, since the corresponding metric is nontrivial \cite{Mor}.
As a result, the later application of the level-$k$ integrability condition \eqref{integrable} requires special care.
We also introduce the additive parameters through $f_i=e^{2\pi iF_i}$, $g_i=e^{2\pi iG_i}$ and $h_i=e^{2\pi iH_i}$.

\begin{table}[t!]
\begin{center}
\begin{tabular}{c|l||c|l}
quantum curves&$(F_{\!1},F_{\!2},G_{\!1},G_{\!2},H_{\!1},H_{\!2})/k$&quantum curves&$(F_{\!1},F_{\!2},G_{\!1},G_{\!2},H_{\!1},H_{\!2})/k$\\\hline
$\widehat{\cal Q}_1^+\widehat{\cal P}_{\rm a}^+\widehat{\cal Q}_2^\pm\widehat{\cal Q}_3^\mp\widehat{\cal P}_{\rm b}^-$&
$(-1,0,-1,-1,1,1)$&
$\widehat{\cal P}_{\rm b}^-\widehat{\cal Q}_2^\pm\widehat{\cal Q}_3^\mp\widehat{\cal P}_{\rm a}^+\widehat{\cal Q}_1^+$&
$(1,0,1,1,-1,-1)$
\\
$\widehat{\cal Q}_2^+\widehat{\cal P}_{\rm a}^+\widehat{\cal Q}_1^\pm\widehat{\cal Q}_3^\mp\widehat{\cal P}_{\rm b}^-$&
$(0,-1,-1,-1,1,1)$&
$\widehat{\cal P}_{\rm b}^-\widehat{\cal Q}_1^\pm\widehat{\cal Q}_3^\mp\widehat{\cal P}_{\rm a}^+\widehat{\cal Q}_2^+$&
$(0,1,1,1,-1,-1)$
\\
$\widehat{\cal Q}_3^+\widehat{\cal P}_{\rm a}^+\widehat{\cal Q}_1^\pm\widehat{\cal Q}_2^\mp\widehat{\cal P}_{\rm b}^-$&
$(1,1,-1,-1,0,0)$&
$\widehat{\cal P}_{\rm b}^-\widehat{\cal Q}_1^\pm\widehat{\cal Q}_2^\mp\widehat{\cal P}_{\rm a}^+\widehat{\cal Q}_3^+$&
$(-1,-1,1,1,0,0)$
\\
$\widehat{\cal Q}_1^+\widehat{\cal P}_{\rm b}^+\widehat{\cal Q}_2^\pm\widehat{\cal Q}_3^\mp\widehat{\cal P}_{\rm a}^-$&
$(-1,0,0,1,0,0)$&
$\widehat{\cal P}_{\rm a}^-\widehat{\cal Q}_2^\pm\widehat{\cal Q}_3^\mp\widehat{\cal P}_{\rm b}^+\widehat{\cal Q}_1^+$&
$(1,0,0,-1,0,0)$
\\
$\widehat{\cal Q}_2^+\widehat{\cal P}_{\rm b}^+\widehat{\cal Q}_1^\pm\widehat{\cal Q}_3^\mp\widehat{\cal P}_{\rm a}^-$&
$(0,-1,0,1,0,0)$&
$\widehat{\cal P}_{\rm a}^-\widehat{\cal Q}_1^\pm\widehat{\cal Q}_3^\mp\widehat{\cal P}_{\rm b}^+\widehat{\cal Q}_2^+$&
$(0,1,0,-1,0,0)$
\\
$\widehat{\cal Q}_3^+\widehat{\cal P}_{\rm b}^+\widehat{\cal Q}_1^\pm\widehat{\cal Q}_2^\mp\widehat{\cal P}_{\rm a}^-$&
$(1,1,0,1,-1,-1)$&
$\widehat{\cal P}_{\rm a}^-\widehat{\cal Q}_1^\pm\widehat{\cal Q}_2^\mp\widehat{\cal P}_{\rm b}^+\widehat{\cal Q}_3^+$&
$(-1,-1,0,-1,1,1)$
\\\hline
$\widehat{\cal P}_{\rm a}^+\widehat{\cal Q}_1^\pm\widehat{\cal Q}_2^\mp\widehat{\cal P}_{\rm b}^-\widehat{\cal Q}_3^+$&
$(0,0,0,-1,0,0)$&
$\widehat{\cal Q}_3^+\widehat{\cal P}_{\rm b}^-\widehat{\cal Q}_1^\pm\widehat{\cal Q}_2^\mp\widehat{\cal P}_{\rm a}^+$&
$(0,0,0,1,0,0)$
\\
$\widehat{\cal P}_{\rm a}^+\widehat{\cal Q}_1^\pm\widehat{\cal Q}_3^\mp\widehat{\cal P}_{\rm b}^-\widehat{\cal Q}_2^+$&
$(0,0,0,-1,0,1)$&
$\widehat{\cal Q}_2^+\widehat{\cal P}_{\rm b}^-\widehat{\cal Q}_1^\pm\widehat{\cal Q}_3^\mp\widehat{\cal P}_{\rm a}^+$&
$(0,0,0,1,0,-1)$
\\
$\widehat{\cal P}_{\rm a}^+\widehat{\cal Q}_2^\pm\widehat{\cal Q}_3^\mp\widehat{\cal P}_{\rm b}^-\widehat{\cal Q}_1^+$&
$(0,0,0,-1,1,0)$&
$\widehat{\cal Q}_1^+\widehat{\cal P}_{\rm b}^-\widehat{\cal Q}_2^\pm\widehat{\cal Q}_3^\mp\widehat{\cal P}_{\rm a}^+$&
$(0,0,0,1,-1,0)$
\\
$\widehat{\cal P}_{\rm b}^+\widehat{\cal Q}_1^\pm\widehat{\cal Q}_2^\mp\widehat{\cal P}_{\rm a}^-\widehat{\cal Q}_3^+$&
$(0,0,1,1,-1,-1)$&
$\widehat{\cal Q}_3^+\widehat{\cal P}_{\rm a}^-\widehat{\cal Q}_1^\pm\widehat{\cal Q}_2^\mp\widehat{\cal P}_{\rm b}^+$&
$(0,0,-1,-1,1,1)$
\\
$\widehat{\cal P}_{\rm b}^+\widehat{\cal Q}_1^\pm\widehat{\cal Q}_3^\mp\widehat{\cal P}_{\rm a}^-\widehat{\cal Q}_2^+$&
$(0,0,1,1,-1,0)$&
$\widehat{\cal Q}_2^+\widehat{\cal P}_{\rm a}^-\widehat{\cal Q}_1^\pm\widehat{\cal Q}_3^\mp\widehat{\cal P}_{\rm b}^+$&
$(0,0,-1,-1,1,0)$
\\
$\widehat{\cal P}_{\rm b}^+\widehat{\cal Q}_2^\pm\widehat{\cal Q}_3^\mp\widehat{\cal P}_{\rm a}^-\widehat{\cal Q}_1^+$&
$(0,0,1,1,0,-1)$&
$\widehat{\cal Q}_1^+\widehat{\cal P}_{\rm a}^-\widehat{\cal Q}_2^\pm\widehat{\cal Q}_3^\mp\widehat{\cal P}_{\rm b}^+$&
$(0,0,-1,-1,0,1)$
\\\hline
$\widehat{\cal Q}_1^\pm\widehat{\cal Q}_2^\mp\widehat{\cal P}_{\rm ab}^-\widehat{\cal Q}_3^+\widehat{\cal P}_{\rm ba}^+$&
$(0,0,-1,0,0,0)$&
$\widehat{\cal P}_{\rm ab}^+\widehat{\cal Q}_3^+\widehat{\cal P}_{\rm ba}^-\widehat{\cal Q}_1^\pm\widehat{\cal Q}_2^\mp$&
$(0,0,1,0,0,0)$
\\
$\widehat{\cal Q}_1^\pm\widehat{\cal Q}_3^\mp\widehat{\cal P}_{\rm ab}^-\widehat{\cal Q}_2^+\widehat{\cal P}_{\rm ba}^+$&
$(0,0,-1,0,0,1)$&
$\widehat{\cal P}_{\rm ab}^+\widehat{\cal Q}_2^+\widehat{\cal P}_{\rm ba}^-\widehat{\cal Q}_1^\pm\widehat{\cal Q}_3^\mp$&
$(0,0,1,0,0,-1)$
\\
$\widehat{\cal Q}_2^\pm\widehat{\cal Q}_3^\mp\widehat{\cal P}_{\rm ab}^-\widehat{\cal Q}_1^+\widehat{\cal P}_{\rm ba}^+$&
$(0,0,-1,0,1,0)$&
$\widehat{\cal P}_{\rm ab}^+\widehat{\cal Q}_1^+\widehat{\cal P}_{\rm ba}^-\widehat{\cal Q}_2^\pm\widehat{\cal Q}_3^\mp$&
$(0,0,1,0,-1,0)$
\\\hline
$\widehat{\cal Q}_1^\pm\widehat{\cal P}_{\rm ab}^-\widehat{\cal Q}_2^+\widehat{\cal P}_{\rm ba}^+\widehat{\cal Q}_3^\mp$&
$(-1,-1,0,0,0,1)$&
$\widehat{\cal Q}_3^\pm\widehat{\cal P}_{\rm ab}^+\widehat{\cal Q}_2^+\widehat{\cal P}_{\rm ba}^-\widehat{\cal Q}_1^\mp$&
$(1,1,0,0,0,-1)$
\\
$\widehat{\cal Q}_1^\pm\widehat{\cal P}_{\rm ab}^-\widehat{\cal Q}_3^+\widehat{\cal P}_{\rm ba}^+\widehat{\cal Q}_2^\mp$&
$(0,1,0,0,-1,0)$&
$\widehat{\cal Q}_2^\pm\widehat{\cal P}_{\rm ab}^+\widehat{\cal Q}_3^+\widehat{\cal P}_{\rm ba}^-\widehat{\cal Q}_1^\mp$&
$(0,-1,0,0,1,0)$
\\
$\widehat{\cal Q}_2^\pm\widehat{\cal P}_{\rm ab}^-\widehat{\cal Q}_1^+\widehat{\cal P}_{\rm ba}^+\widehat{\cal Q}_3^\mp$&
$(-1,-1,0,0,1,0)$&
$\widehat{\cal Q}_3^\pm\widehat{\cal P}_{\rm ab}^+\widehat{\cal Q}_1^+\widehat{\cal P}_{\rm ba}^-\widehat{\cal Q}_2^\mp$&
$(1,1,0,0,-1,0)$
\\
$\widehat{\cal Q}_2^\pm\widehat{\cal P}_{\rm ab}^-\widehat{\cal Q}_3^+\widehat{\cal P}_{\rm ba}^+\widehat{\cal Q}_1^\mp$&
$(1,0,0,0,0,-1)$&
$\widehat{\cal Q}_1^\pm\widehat{\cal P}_{\rm ab}^+\widehat{\cal Q}_3^+\widehat{\cal P}_{\rm ba}^-\widehat{\cal Q}_2^\mp$&
$(-1,0,0,0,0,1)$
\\
$\widehat{\cal Q}_3^\pm\widehat{\cal P}_{\rm ab}^-\widehat{\cal Q}_1^+\widehat{\cal P}_{\rm ba}^+\widehat{\cal Q}_2^\mp$&
$(0,1,0,0,0,0)$&
$\widehat{\cal Q}_2^\pm\widehat{\cal P}_{\rm ab}^+\widehat{\cal Q}_1^+\widehat{\cal P}_{\rm ba}^-\widehat{\cal Q}_3^\mp$&
$(0,-1,0,0,0,0)$
\\
$\widehat{\cal Q}_3^\pm\widehat{\cal P}_{\rm ab}^-\widehat{\cal Q}_2^+\widehat{\cal P}_{\rm ba}^+\widehat{\cal Q}_1^\mp$&
$(1,0,0,0,0,0)$&
$\widehat{\cal Q}_1^\pm\widehat{\cal P}_{\rm ab}^+\widehat{\cal Q}_2^+\widehat{\cal P}_{\rm ba}^-\widehat{\cal Q}_3^\mp$&
$(-1,0,0,0,0,0)$
\\\hline
$\widehat{\cal P}_{\rm ab}^-\widehat{\cal Q}_1^+\widehat{\cal P}_{\rm ba}^+\widehat{\cal Q}_2^\pm\widehat{\cal Q}_3^\mp$&
$(-1,0,1,0,0,0)$&
$\widehat{\cal Q}_2^\pm\widehat{\cal Q}_3^\mp\widehat{\cal P}_{\rm ab}^+\widehat{\cal Q}_1^+\widehat{\cal P}_{\rm ba}^-$&
$(1,0,-1,0,0,0)$
\\
$\widehat{\cal P}_{\rm ab}^-\widehat{\cal Q}_2^+\widehat{\cal P}_{\rm ba}^+\widehat{\cal Q}_1^\pm\widehat{\cal Q}_3^\mp$&
$(0,-1,1,0,0,0)$&
$\widehat{\cal Q}_1^\pm\widehat{\cal Q}_3^\mp\widehat{\cal P}_{\rm ab}^+\widehat{\cal Q}_2^+\widehat{\cal P}_{\rm ba}^-$&
$(0,1,-1,0,0,0)$
\\
$\widehat{\cal P}_{\rm ab}^-\widehat{\cal Q}_3^+\widehat{\cal P}_{\rm ba}^+\widehat{\cal Q}_1^\pm\widehat{\cal Q}_2^\mp$&
$(1,1,1,0,-1,-1)$&
$\widehat{\cal Q}_1^\pm\widehat{\cal Q}_2^\mp\widehat{\cal P}_{\rm ab}^+\widehat{\cal Q}_3^+\widehat{\cal P}_{\rm ba}^-$&
$(-1,-1,-1,0,1,1)$
\end{tabular}
\end{center}
\caption{Factorized quantum curves identified for vertices of $E_6$.
For vertices that admit multiple realizations, we list multiple equivalent realizations whenever possible.
The vertices in the left column, together with $k(1,0,0,0,-1,0)$, $k(0,1,0,0,0,-1)$ and $k(-1,-1,0,0,1,1)$ in \eqref{E6unid}, form the chiral representation ${\bf 27}$ of $E_6$. Likewise, the vertices in the right column, together with $k(-1,0,0,0,1,0)$, $k(0,-1,0,0,0,1)$ and $k(1,1,0,0,-1,-1)$ form the conjugate representation $\overline{\bf 27}$.}
\label{E6vertex}
\end{table}

Although a complete interpretation in terms of brane configurations is still lacking, the rectangular support of the $E_6$ quantum curve already provides a useful clue.
Indeed, the Newton polygon is a $2\times 3$ rectangle, naturally suggesting a configuration with two $(1,k)$5-branes and three NS5-branes, while the shifts of powers are naturally interpreted as effects of FI parameters (or mass deformations).

Despite the absence of a complete brane interpretation, we can nevertheless generate all facets by acting with the $E_6$ Weyl group on the level-$k$ integrability condition \eqref{integrable},
\begin{align}
G_2\le k,
\end{align}
which follows from a careful treatment of the nontrivial metric.
There are $72$ facets in total, corresponding to the 72 roots of $E_6$.
Again, these facets can be regarded as the constraints \eqref{voronoi} that the origin is closer than to any other $E_6$ roots.
Hence, the fundamental domain is the Voronoi polytope of the $E_6$ root lattice. 
Then, in principle, the vertices are obtained by solving for the intersections of these facets.
Although determining all vertices explicitly is tedious in six dimensions, the $E_6$ symmetry suggests that the vertices should be organized into Weyl orbits of $E_6$.
Guided by this expectation, we find that the vertices are consistent with the Weyl orbits of the weights of the minuscule representations $\bm{27}$ and $\overline{\bm{27}}$.
Although all vertices belong to the Weyl orbits of the minuscule representations for the $D_5$ and $E_6$ cases considered so far, this is not a general feature.
In appendix \ref{rootpolytope}, we discuss the dual description in terms of root polytopes of exceptional algebras and explain the expected structure of vertices.
This becomes relevant for the $E_7$ case studied below.

Here, as previously, we attempt to identify these vertices with factorized quantum curves with $\widehat{\cal Q}^\pm$ and $\widehat{\cal P}^\pm$.
To reproduce the asymptotic degeneracy in \eqref{e6curve} which indicates that the two asymptotic values associated with $g_1$ must differ by a factor of $q$, we need to consider factorized quantum curves with exactly one $\widehat Q$ between two $\widehat{\cal P}^\pm$.
For this reason, the operators inserted between them can only be $\widehat{\cal Q}^+$, $\widehat{\cal Q}^+\widehat{\cal Q}^-$ or $\widehat{\cal Q}^-\widehat{\cal Q}^+$ in terms of $\widehat{\cal Q}^\pm$.
As in the $D_5$ case, almost all vertices of the fundamental domain are reproduced by factorized quantum curves built from $\widehat{\cal Q}^\pm$ and $\widehat{\cal P}^\pm$, in a manner consistent with the degeneracies of the $E_6$ curve. 
The results are given in table \ref{E6vertex}.
Here we label the canonical operators $\widehat{\cal Q}^\pm$ and $\widehat{\cal P}^\pm$ by $1,2,3$ and $\text{a},\text{b}$ respectively.

In \eqref{boundary} we have seen that only boundary terms have a clean linear interpretation in terms of relative ranks.
Remarkably, the factorized spectral operators constructed from $\widehat{\cal Q}^\pm$ and $\widehat{\cal P}^\pm$ in table \ref{E6vertex} also correctly reproduce the nonlinear coefficient of the interior $\widehat Q$ term in \eqref{e6curve},
\begin{align}
g_1(f_1+f_2+f_3)+h_1^{-1}+h_2^{-1}+h_3^{-1},
\label{E6int}
\end{align}
which is known to be completely determined by the boundary terms for the $E_6$ quantum curve.
Thus, the factorized curves reproduce not only the boundary data but also the interior structure of the $E_6$ curve.
Although the factorized quantum curves were constructed merely to reproduce the degeneracy pattern of asymptotic values and do not manifest the $E_6$ Weyl-group symmetry a priori, the agreement of the interior coefficient implies that the resulting curve coincides with the $E_6$ quantum curve itself and hence inherits the full $E_6$ Weyl-group symmetry.
Within our realization of the vertices of the fundamental domain in terms of factorized quantum curves, this is a highly nontrivial observation, which strongly supports our proposal.

As in the case of the $D_5$ curve, there are six unidentified parameters.
\begin{align}
(\pm k,0,0,0,\mp k,0),\quad
(0,\pm k,0,0,0,\mp k),\quad
(\mp k,\mp k,0,0,\pm k,\pm k).
\label{E6unid}
\end{align}
As in the $D_5$ case, these may be interpreted as configurations in which the 5-brane charges are effectively averaged, leaving only nontrivial FI parameters.

\subsection{$E_7$}

Next, we repeat the same analysis for the $E_7$ quantum curve.
The $E_7$ quantum curve in the rectangular realization $\widehat P^m\widehat Q^n$ ($-1\le m\le 1,-2\le n\le 2$) is given by
\begin{align}
&\widehat H/\alpha=q^{-1}\widehat Q^2(\widehat P+q^{\frac{3}{2}}g_1)(\widehat P+q^{\frac{1}{2}}g_1)\widehat P^{-1}\nonumber\\
&\;+q^{-\frac{1}{2}}\widehat Q(\widehat P+q^{\frac{1}{2}}g_1)((f_1+f_2+f_3+f_4)\widehat P+q^{\frac{1}{2}}(h_1^{-1}+h_2^{-1}+h_3^{-1}+h_4^{-1}))\widehat P^{-1}\nonumber\\
&\;+(f_1f_2+\cdots+f_3f_4)\widehat P+E/\alpha+(h_1^{-1}h_2^{-1}+\cdots+h_3^{-1}h_4^{-1})\widehat P^{-1}\nonumber\\
&\;+q^{\frac{1}{2}}f_1f_2f_3f_4\widehat Q^{-1}(\widehat P+q^{-\frac{1}{2}}g_2)((f_1^{-1}+f_2^{-1}+f_3^{-1}+f_4^{-1})\widehat P+q^{-\frac{1}{2}}g_1g_2(h_1+h_2+h_3+h_4))\widehat P^{-1}\nonumber\\
&\;+qf_1f_2f_3f_4\widehat Q^{-2}(\widehat P+q^{-\frac{3}{2}}g_2)(\widehat P+q^{-\frac{1}{2}}g_2)\widehat P^{-1},
\label{e7curve}
\end{align}
with $f_1f_2f_3f_4(g_1g_2)^2h_1h_2h_3h_4=1$ in the notation of \cite{Mor}.
\begin{figure}[!t]
\vspace{-12mm}
\centering\includegraphics[width=16cm]{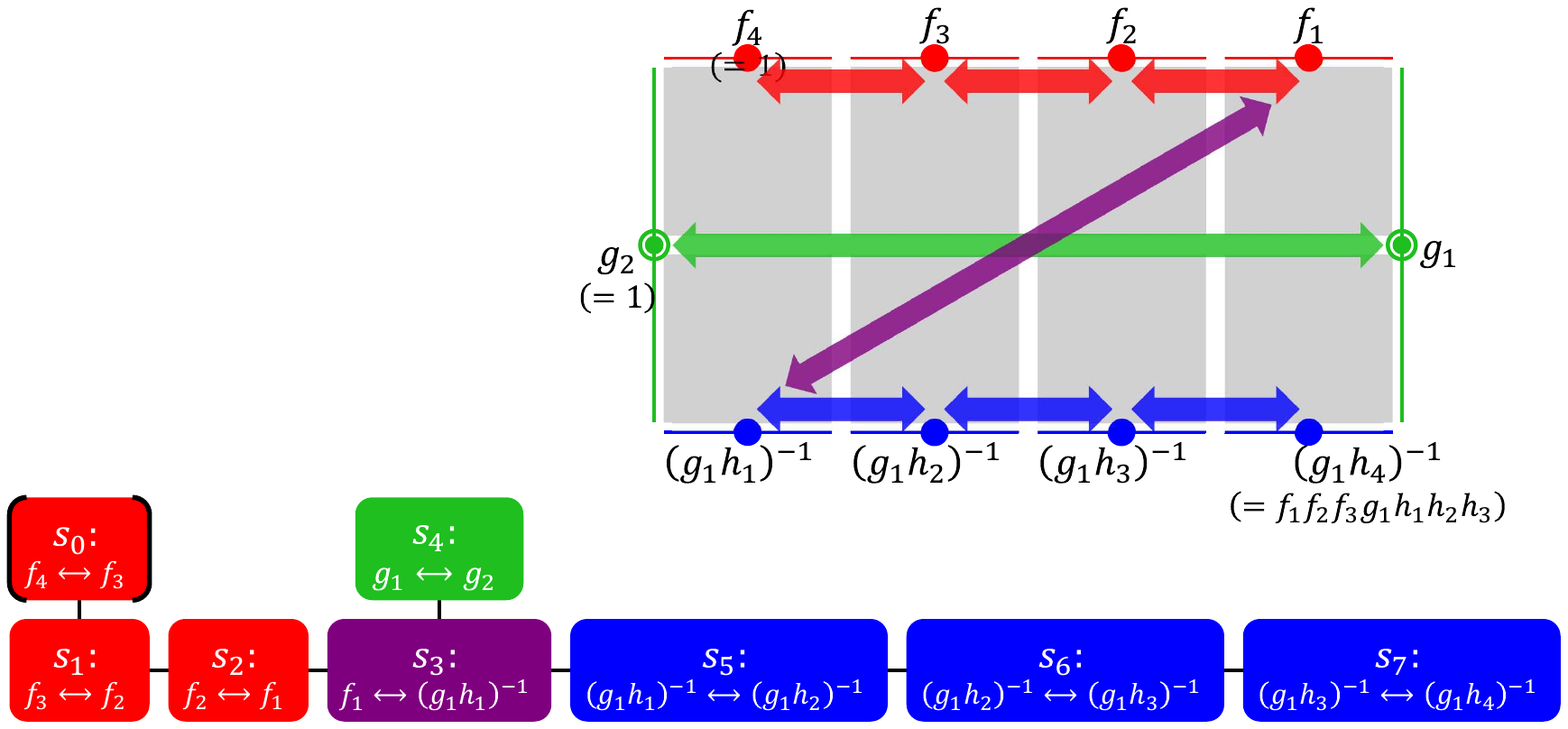}
\vspace{-24mm}
\caption{Asymptotic values and Weyl reflections for the $E_7$ quantum curve.}
\label{e7fig}
\end{figure}
The asymptotic values and the Weyl reflections for this curve are given in figure \ref{e7fig}.
The Newton polygon is a $2\times 4$ rectangle, corresponding to a configuration with two $(1,k)$5-branes and four NS5-branes in the presence of FI parameters.
We often fix the gauge ambiguity in the parameter space by setting
\begin{align}
f_4=g_2=1,\quad h_4=(f_1f_2f_3g_1^2h_1h_2h_3)^{-1}.
\end{align}
Then, the Weyl reflections, simple roots, fundamental weights and metric are obtained straightforwardly \cite{Mor}.

The fundamental domain in the ${\cal H}$-description is obtained by applying the $E_7$ Weyl group to the level-$k$ integrability condition \eqref{integrable},
\begin{align}
F_3\le k.
\end{align}
There are $126$ facets in total, corresponding to the $126$ roots of $E_7$.
As shown in \cite{FMMN}, the resulting inequalities are consistent with the condition for the termination of duality cascades, using a brane realization consisting of two $(1,k)$5-branes and four NS5-branes arranged as NS5-NS5-$(1,k)$5-NS5-NS5-$(1,k)$5, which reproduces the degeneracies of the $E_7$ quantum curve.
Then, guided by the $E_7$ symmetry, we identify among the vertices a Weyl orbit corresponding to the minuscule representation $\bm{56}$.
See appendix \ref{rootpolytope} for a discussion of another Weyl orbit that also appears among the vertices of the fundamental domain.

\begin{table}[!t]
\begin{center}
\begin{tabular}{r|lll}
quantum curves&\multicolumn{3}{l}{$(F_1,F_2,F_3,G_1,H_1,H_2,H_3)/k$}\\\hline\hline
$\widehat{\cal Q}^\pm\widehat{\cal Q}^\mp\widehat{\cal P}^+\widehat{\cal Q}^\pm\widehat{\cal Q}^\mp\widehat{\cal P}^-$
&$(0,-1,-1,-1,1,1,1)$,&\!\!\!$(-1,0,-1,-1,1,1,1)$,&\!\!\!$(-1,-1,0,-1,1,1,1)$,\\
&$(0,1,1,-1,0,0,0)$,&\!\!\!$(1,0,1,-1,0,0,0)$,&\!\!\!$(1,1,0,-1,0,0,0)$\\\hline
$\widehat{\cal Q}^\pm\widehat{\cal Q}^\mp\widehat{\cal P}^-\widehat{\cal Q}^\pm\widehat{\cal Q}^\mp\widehat{\cal P}^+$
&$(0,0,0,-1,1,0,0)$,&\!\!\!$(0,0,0,-1,0,1,0)$,&\!\!\!$(0,0,0,-1,0,0,1)$,\\
&$(0,0,0,-1,0,1,1)$,&\!\!\!$(0,0,0,-1,1,0,1)$,&\!\!\!$(0,0,0,-1,1,1,0)$\\\hline
$\widehat{\cal Q}^\mp\widehat{\cal P}^+\widehat{\cal Q}^\pm\widehat{\cal Q}^\mp\widehat{\cal P}^-\widehat{\cal Q}^\pm$
&$(-1,0,0,0,0,1,0)$,&\!\!\!$(-1,0,0,0,0,0,1)$,&\!\!\!$(-1,0,0,0,0,0,0)$,\\
&$(0,-1,0,0,1,0,0)$,&\!\!\!$(0,-1,0,0,0,0,1)$,&\!\!\!$(0,-1,0,0,0,0,0)$,\\
&$(0,0,-1,0,1,0,0)$,&\!\!\!$(0,0,-1,0,0,1,0)$,&\!\!\!$(0,0,-1,0,0,0,0)$,\\
&$(1,1,1,0,-1,-1,0)$,&\!\!\!$(1,1,1,0,-1,0,-1)$,&\!\!\!$(1,1,1,0,0,-1,-1)$\\\hline
$\widehat{\cal Q}^\mp\widehat{\cal P}^-\widehat{\cal Q}^\pm\widehat{\cal Q}^\mp\widehat{\cal P}^+\widehat{\cal Q}^\pm$
&$(-1,-1,-1,0,0,1,1)$,&\!\!\!$(-1,-1,-1,0,1,0,1)$,&\!\!\!$(-1,-1,-1,0,1,1,0)$,\\
&$(0,0,1,0,0,0,0)$,&\!\!\!$(0,0,1,0,0,-1,0)$,&\!\!\!$(0,0,1,0,-1,0,0)$,\\
&$(0,1,0,0,0,0,0)$,&\!\!\!$(0,1,0,0,0,0,-1)$,&\!\!\!$(0,1,0,0,-1,0,0)$,\\
&$(1,0,0,0,0,0,0)$,&\!\!\!$(1,0,0,0,0,0,-1)$,&\!\!\!$(1,0,0,0,0,-1,0)$\\\hline
$\widehat{\cal P}^+\widehat{\cal Q}^\pm\widehat{\cal Q}^\mp\widehat{\cal P}^-\widehat{\cal Q}^\pm\widehat{\cal Q}^\mp$
&$(0,0,0,1,-1,-1,0)$,&\!\!\!$(0,0,0,1,-1,0,-1)$,&\!\!\!$(0,0,0,1,0,-1,-1)$,\\
&$(0,0,0,1,0,0,-1)$,&\!\!\!$(0,0,0,1,0,-1,0)$,&\!\!\!$(0,0,0,1,-1,0,0)$\\\hline
$\widehat{\cal P}^-\widehat{\cal Q}^\pm\widehat{\cal Q}^\mp\widehat{\cal P}^+\widehat{\cal Q}^\pm\widehat{\cal Q}^\mp$
&$(-1,-1,0,1,0,0,0)$,&\!\!\!$(-1,0,-1,1,0,0,0)$,&\!\!\!$(0,-1,-1,1,0,0,0)$,\\
&$(1,1,0,1,-1,-1,-1)$,&\!\!\!$(1,0,1,1,-1,-1,-1)$,&\!\!\!$(0,1,1,1,-1,-1,-1)$
\end{tabular}
\end{center}
\caption{Factorized quantum curves identified for vertices of $E_7$.}
\label{E7vertex}
\end{table}

Again, to identify these vertices with factorized quantum curves, let us study the factorized quantum curves constructed by $\widehat{\cal Q}^\pm$ and $\widehat{\cal P}^\pm$.
To match the degeneracies of the $E_7$ quantum curve \eqref{e7curve}, where the pairs of asymptotic values $-q^{\frac{3}{2}}g_1$, $-q^{\frac{1}{2}}g_1$ and $-q^{-\frac{3}{2}}g_2$, $-q^{-\frac{1}{2}}g_2$ differ by factors of $q$, we need to consider orderings in which both $\widehat Q$ and $\widehat Q^{-1}$ appear between two $\widehat{\cal P}^\pm$.
For this reason, between them we can only have either $\widehat{\cal Q}^+\widehat{\cal Q}^-$ or $\widehat{\cal Q}^-\widehat{\cal Q}^+$ in terms of $\widehat{\cal Q}^\pm$.
The results are listed in table \ref{E7vertex}.
We find again that the asymptotic values correctly reproduce the $E_7$ minuscule representation ${\bf 56}$ obtained from the inequalities in the ${\cal H}$-description.
Again the factorized quantum curves also reproduce correctly the coefficients of the internal terms $\widehat Q$ and $\widehat Q^{-1}$ in \eqref{e7curve},
\begin{align}
&g_1(f_1+f_2+f_3+f_4)+h_1^{-1}+h_2^{-1}+h_3^{-1}+h_4^{-1},\nonumber\\
&(f_1f_2f_3f_4g_2)(f_1^{-1}+f_2^{-1}+f_3^{-1}+f_4^{-1}+g_1(h_1+h_2+h_3+h_4)).
\label{E7int}
\end{align}
This is a nontrivial check that the vertices of the fundamental domain are realized by factorized quantum curves constructed from $\widehat{\cal Q}^\pm$ and $\widehat{\cal P}^\pm$.

As in the $D_5$ and $E_6$ cases, there are eight unidentified parameters
\begin{align}
&(\pm k,0,0,0,\mp k,0,0),\quad
(0,\pm k,0,0,0,\mp k,0),\quad
(0,0,\pm k,0,0,0,\mp k),\nonumber\\
&(\mp k,\mp k,\mp k,0,\pm k,\pm k,\pm k),
\label{E7unid}
\end{align}
which may be interpreted as configurations in which the 5-brane charges are effectively averaged, leaving only nontrivial FI parameters.

\section{Conclusion and discussion}

Duality cascades are identified as discrete translations in the parameter space of relative ranks.
For this reason, the questions of whether duality cascades terminate in finitely many steps and whether their endpoints are unique can be translated into the question of whether the fundamental domain is a parallelotope, which tiles the whole parameter space by discrete translations.
This question was studied in \cite{FMS,MO} and the fact that the vertices of the polytope correspond to brane configurations with vanishing relative ranks plays an essential role.
As a first step toward studying quantum curves associated with del Pezzo geometries with FI parameters \cite{FMMN}, it is important to understand the interpretation of the vertices.

In this paper, we introduce the quantum operators $\widehat{\cal Q}^\pm$ and $\widehat{\cal P}^\pm$ \eqref{QPpm} and find that the vertices of the fundamental domain belonging to the minuscule representations are naturally realized as factorized quantum curves built from these operators.
From this perspective, the operators $\widehat{\cal Q}^\pm$ and $\widehat{\cal P}^\pm$ serve as fundamental building blocks of the factorized quantum curves at the minuscule vertices of the fundamental domain and correspond to ``extremal'' 5-branes dressed with FI parameters.
Our results provide a first step towards the realization of the ${\cal V}$-description for these exceptional quantum curves, while suggesting that the class of factorized quantum curves considered here naturally singles out the minuscule vertices of the Voronoi polytope.
This suggests that the physically distinguished objects are not the operator monomials appearing in an expanded quantum curve, but rather the elementary operator factors $\widehat{\cal Q}^\pm$ and $\widehat{\cal P}^\pm$ from which the factorized quantum curve is built.
Below we list several directions for future investigations.

First, from the viewpoint of the factorized quantum curves, it is important to understand the physical characterization of these ``extremal'' 5-branes.
It is somewhat remarkable that the FI parameters participate on an equal footing with the relative ranks in defining the affine Weyl chamber.
Correspondingly, the operators $\widehat{\cal Q}$ and $\widehat{\cal P}$ \eqref{QPhalf} associated with the original 5-branes are refined into the extremal operators $\widehat{\cal Q}^\pm$ and $\widehat{\cal P}^\pm$ \eqref{QPpm} in the presence of FI parameters.
It would be interesting to understand the physical significance of this refinement directly in the brane picture.
It may also be related to the convergence range of relative ranks appearing in the computation of the partition functions \cite{HMO2,MN9}, and understanding this connection could shed light on the nature of the vertices.
Furthermore, even for the minuscule vertices of the $D_5$, $E_6$ and $E_7$ quantum curves studied in this paper, there remain unidentified configurations in \eqref{D5unid}, \eqref{E6unid} and \eqref{E7unid}.
Despite the interpretation suggested above, it would be interesting to construct the curves explicitly in terms of $\widehat{\cal Q}^\pm$ and $\widehat{\cal P}^\pm$ and see their physical interpretation.
Also, although the $E_7$ fundamental domain contains another Weyl orbit of vertices of length $576$, which does not arise from a minuscule representation, its realization in terms of quantum curves remains unclear.

Second, of course, it is desirable to study $E_8$ as well.
The main difficulty is that the degeneracy is not consistent with the proposed factorized quantum curves.
For the $E_6$ and $E_7$ quantum curves, it is known \cite{Mor,MY} that the degeneracies are split by the powers of $q$, (see \eqref{e6curve} and \eqref{e7curve}).
This is also the case for the $E_8$ quantum curve \cite{Mor,MY}.
For the rectangular realization of the $E_8$ quantum curve $\widehat P^m\widehat Q^n$ ($-1\le m\le 2,-3\le n\le 3$), we need to construct the curve using two $\widehat{\cal P}^+$ and one $\widehat{\cal P}^-$ along with three pairs of $\widehat{\cal Q}^+$ and $\widehat{\cal Q}^-$.
The degeneracy condition for the right and left sides requires that exactly one $\widehat{\cal Q}^+$ and one $\widehat{\cal Q}^-$ be placed between two $\widehat{\cal P}^\pm$ operators, forming $\widehat{\cal P}^\pm\widehat{\cal Q}^+\widehat{\cal Q}^-\widehat{\cal P}^\pm\widehat{\cal Q}^+\widehat{\cal Q}^-\widehat{\cal P}^\pm$, while the degeneracy condition for the upper side requires that a $\widehat{\cal P}^+$ operator be inserted between three pairs of $\widehat{\cal Q}^\pm$ operators.
These conditions are mutually incompatible.
It is tempting to speculate that the obstruction to realizing the vertices with factorized quantum curves for $E_8$ is related to the absence of minuscule representations for $E_8$, in contrast to the $D_5$, $E_6$ and $E_7$ cases.
This speculation is further supported by the $E_7$ case, where the Weyl orbit of length $576$ also lies outside the minuscule representations and likewise lacks a realization in terms of factorized quantum curves.
The resolution may require a further refinement of the canonical operators $\widehat{\cal Q}^\pm$ and $\widehat{\cal P}^\pm$ in \eqref{QPpm} or a novel extension of the factorization structure studied in this paper.
We would like to seek such a resolution or an alternative interpretation of the $E_8$ quantum curve.

Third, to reproduce the degeneracies of the $E_6$ and $E_7$ quantum curves, we need to consider spectral operators built from $\widehat{\cal Q}^\pm$ and $\widehat{\cal P}^\pm$ in orderings satisfying certain constraints.
The role of operators that do not satisfy these constraints remains unclear.
One natural possibility is that, after relaxing the degeneracy constraints and thereby generalizing the quantum curves, factorized curves constructed from $\widehat{\cal Q}^\pm$ and $\widehat{\cal P}^\pm$ in arbitrary orderings may still correspond to the vertices of a generalized fundamental domain.
We would like to clarify their role.

It is nontrivial that the coefficients of the internal lattice points of the Newton polygons are also correctly reproduced from the factorized quantum curves, as in \eqref{E6int} and \eqref{E7int}.
The constant term $E$ in \eqref{d5curve}, \eqref{e6curve} and \eqref{e7curve} is perhaps even more intriguing.
Since this term is invariant under the Weyl group, it often plays no essential role in discussions based on Weyl orbits and is therefore usually left implicit.
Nevertheless, factorization requires the constant term $E$ to take specific values.
We observe that, in the Weyl ordering, vertices belonging to the same representation share the same value of $E$.
It would be interesting to understand the significance of this observation.

From the viewpoint of Painlev\'e systems, the factorization studied in this paper may be related to special solutions of Painlev\'e equations, such as hypergeometric solutions \cite{KNY}.
It is also known that factorization of quantum operators can reveal hidden modular or mock modular structures \cite{Tsu}.
In this respect, it would be interesting to clarify whether the factorized quantum curves at the vertices of the fundamental domain characterize distinguished sectors for solutions of $q$-Painlev\'e equations.
We hope that our results provide useful insights into the structure underlying the $D_5$, $E_6$ and $E_7$ $q$-Painlev\'e equations.

From a mathematical perspective, our main observation is that the minuscule vertices of the Voronoi polytopes of the $D_5$, $E_6$ and $E_7$ root lattices are naturally associated with factorized quantum curves constructed from the canonical operators $\widehat{\cal Q}^\pm$ and $\widehat{\cal P}^\pm$.
Although a small number of vertices remain unidentified (\eqref{D5unid}, \eqref{E6unid}, \eqref{E7unid}), we provide extensive evidence for this correspondence by explicitly enumerating all cases.
It would be interesting to find a conceptual proof and clarify the underlying mathematical structure behind this correspondence.

Del Pezzo geometries play a central role in the study of five-dimensional gauge theories with exceptional flavor symmetries \cite{Sei}.
Since the quantum curves studied in this paper can be regarded as quantizations of the corresponding del Pezzo geometries \cite{BBT,KY,CL}, it would be interesting to understand whether the factorized quantum curves and the realization of the vertices found in this paper admit a natural interpretation in this five-dimensional framework.

The noncommutative structures appearing in our study of quantum curves and quantum tori are reminiscent of the wall-crossing structures appearing in \cite{KoSo}.
It would be interesting to clarify possible underlying relations.

\appendix

\section{Root polytopes}\label{rootpolytope}

In this appendix, we shall summarize some results on the root polytopes associated with the $E_6$, $E_7$ and $E_8$ root systems and relate them to the fundamental domains studied in the main text.
We shall see that the polytopes appearing as the fundamental domains of duality cascades can be regarded as dual polytopes of these root polytopes.

Root polytopes associated with Lie algebras are defined as the convex hulls of their roots.
For $E_6$, $E_7$ and $E_8$ algebras, these correspond to famous Gosset polytopes $1_{22}$, $2_{31}$ and $4_{21}$, whose numbers of vertices and facets \cite{wiki} are shown in table \ref{Gosset}.
The Gosset symbols $1_{22}$, $2_{31}$ and $4_{21}$ reflect the shapes of the corresponding Dynkin diagrams.
The integers $i$, $j$ and $k$ in $k_{ij}$ roughly encode the lengths of the branches measured from the bifurcation node together with the choice of the distinguished node.

\begin{table}
\begin{center}
\begin{tabular}{c|c||c|c}
Roots&Gosset&vertices&facets\\\hline
$E_6$&$1_{22}$&$72$&$54=27+27$\\
$E_7$&$2_{31}$&$126$&$632=56+576$\\
$E_8$&$4_{21}$&$240$&$19440=2160+17280$
\end{tabular}
\end{center}
\caption{Numbers of facets for the $E_6$, $E_7$ and $E_8$ root polytopes.}
\label{Gosset}
\end{table} 

\begin{figure}[!t]
\centering
\includegraphics[width=12cm]{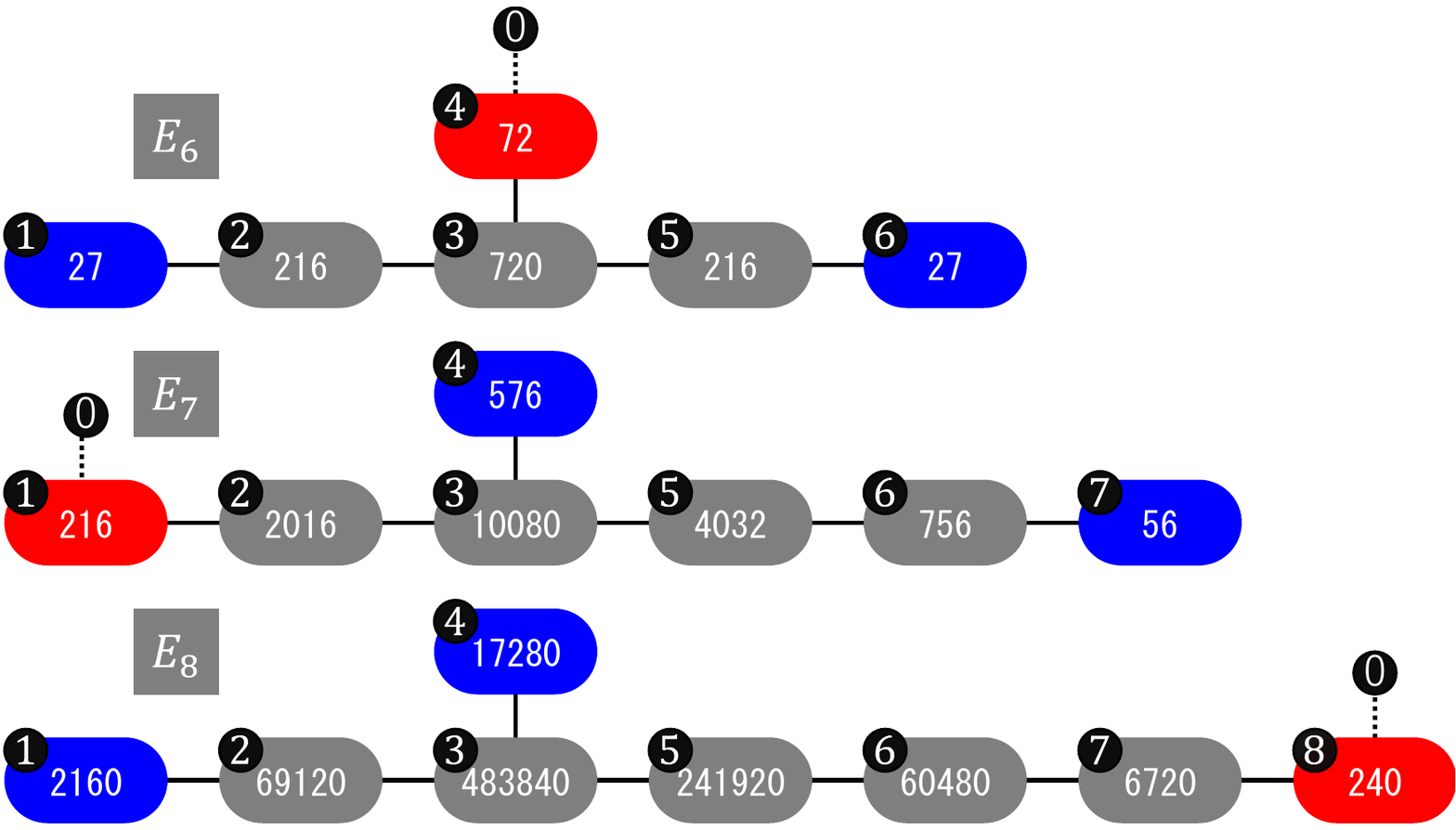}
\vspace{-10mm}
\caption{Length of Weyl orbit for each fundamental weight.
The length described in each node is obtained by dividing the order of the original Weyl group by that of the stabilizer subgroup, which is read off by removing the corresponding node.
The vertices correspond to the fundamental weights in the adjoint representations (red), while the facets correspond to those in the other two ends of the Dynkin diagram (blue).
These orbit lengths are closely related to the representations associated with the corresponding fundamental weights:
Indeed, two numbers coincide for the minuscule representations, while they differ for the others because the Weyl orbit does not exhaust all weights of the representation.
}
\label{orbits}
\end{figure}

Each summand in the decomposition of the facet counts in table \ref{Gosset} corresponds to a set of facets belonging to the same Weyl orbit.
Namely, each facet number is the length of a Weyl orbit obtained by quotienting the Weyl group by the corresponding stabilizer subgroup, which is read off from the Dynkin diagram by removing the associated node.
Rewriting these decompositions as
\begin{align}
27+27&=|W(E_6)|(|W(D_5)|^{-1}+|W(D_5)|^{-1}),\nonumber\\
56+576&=|W(E_7)|(|W(E_6)|^{-1}+|W(A_6)|^{-1}),\nonumber\\
2160+17280&=|W(E_8)|(|W(D_7)|^{-1}+|W(A_7)|^{-1}),
\end{align}
reveals the corresponding stabilizer subgroups, and hence the associated fundamental weights, directly from the denominators.
This shows that the vertices of the root polytope, namely the roots in the adjoint representation, are associated with the node adjacent to the affine node, while the facets are associated with the two terminal nodes connected to the adjoint node through the bifurcation.
More generally, the length of the Weyl orbit for each fundamental weight is given in figure \ref{orbits}.

The polytope representing the fundamental domain studied in the main text is dual to the root polytope introduced above.
Let $R$ denote the set of roots.
The root polytope is defined as the convex hull of $R$, while the fundamental domain is the intersection of the inequalities $\bm\alpha\cdot\bm\lambda\le 1$ for all $\bm\alpha\in R$.
These two constructions are dual to each other.
In particular, determining the facets of the root polytope and determining the vertices of the fundamental domain reduce to essentially the same problem.
In both cases, one seeks a vector $\bm\lambda$ such that $\bm\alpha\cdot\bm\lambda\le 1$ for all $\bm\alpha\in R$, and the roots saturating the equality $\bm\alpha\cdot\bm\lambda=1$ span the corresponding facet.
Thus, the two polytopes are dual to each other, with vertices and facets exchanged.

This suggests that the vertices of the fundamental domain are naturally associated with the facets of the root polytope.
Although, as emphasized in the main text, we are unable to generate all vertices directly from the inequalities because of computational limitations, we can nevertheless test this expectation explicitly.
After generating the full root system defining the inequalities, it is natural to look for highly symmetric vertices associated with the fundamental weights.
For a point proportional to a fundamental weight, the inequalities determine a finite interval for the proportionality constant within which the point remains inside the polytope.
The endpoints of this interval lie on the boundary, and in some cases they satisfy sufficiently many independent equalities to become vertices.
Since the highest root is expanded in terms of the simple roots with coefficients given by the marks $a_i$, the rescaled fundamental weights $\lambda_i/a_i$ automatically saturate the inequality associated with the highest root.
They are therefore natural candidates for vertices lying on the boundary of the fundamental domain.

Namely, for $E_6$, the rescaled fundamental weights
\begin{align}
\lambda_1,\quad\lambda_2/2,\quad\lambda_3/3,\quad\lambda_4/2,\quad\lambda_5/2,\quad\lambda_6,
\end{align}
are located on the boundary of the fundamental domain, and only $\lambda_1$ and $\lambda_6$ saturate sufficiently many independent equalities to become vertices of the polytope.
For $E_7$, the rescaled fundamental weights
\begin{align}
\lambda_1/2,\quad\lambda_2/3,\quad\lambda_3/4,\quad\lambda_4/2,\quad\lambda_5/3,\quad\lambda_6/2,\quad\lambda_7,
\end{align}
are located on the boundary of the fundamental domain, and only $\lambda_4/2$ and $\lambda_7$ saturate sufficiently many independent equalities to become vertices of the polytope.
Although not used in the main text, for $E_8$, the rescaled fundamental weights
\begin{align}
\lambda_1/2,\quad\lambda_2/4,\quad\lambda_3/6,\quad\lambda_4/3,\quad\lambda_5/5,\quad\lambda_6/4,\quad\lambda_7/3,\quad\lambda_8/2,
\end{align}
are located on the boundary of the fundamental domain, and only $\lambda_1/2$ and $\lambda_4/3$ saturate sufficiently many independent equalities to become vertices of the polytope.
Just as for the facets of the root polytopes, these fundamental weights specifying the vertices are associated with the two terminal nodes of the Dynkin diagram, which are connected to the adjoint node through the bifurcation.
Their Weyl orbits contain exactly as many vertices as facets of the corresponding root polytopes.
The Weyl orbits of the minuscule weights, namely $\lambda_1$ and $\lambda_6$ for $E_6$ and $\lambda_7$ for $E_7$, are precisely those realized by the factorized quantum curves constructed in the main text.
In general, however, the factorized quantum curves do not exhaust all vertices of the fundamental domain. 

\section{Simple roots and fundamental weights}\label{rootweight}

In this appendix, we record the simple roots and fundamental weights read off from the Weyl reflections for completeness \cite{Mor}.
For $E_6$ the simple roots and the fundamental weights are given by
\begin{align}
\alpha_1&=(-1,-2,0,0,1,1)^\text{T},&
\omega_1&=(-1,-1,1,1,0,0)^\text{T},\nonumber\\
\alpha_2&=(-1,1,0,0,0,0)^\text{T},&
\omega_2&=(-1,0,2,2,-1,-1)^\text{T},\nonumber\\
\alpha_3&=(1,0,1,0,0,-1)^\text{T},&
\omega_3&=(0,0,3,3,-2,-2)^\text{T},\nonumber\\
\alpha_4&=(0,0,-1,1,0,0)^\text{T},&
\omega_4&=(0,0,1,2,-1,-1)^\text{T},\nonumber\\
\alpha_5&=(0,0,0,0,-1,1)^\text{T},&
\omega_5&=(0,0,2,2,-2,-1)^\text{T},\nonumber\\
\alpha_6&=(0,0,0,0,0,-1)^\text{T},&
\omega_6&=(0,0,1,1,-1,-1)^\text{T}.
\end{align}
For $E_7$ they are
\begin{align}
\alpha_1&=(0,-1,1,0,0,0,0)^\text{T},&
\omega_1&=(1,1,2,0,-1,-1,-1)^\text{T},\nonumber\\
\alpha_2&=(-1,1,0,0,0,0,0)^\text{T},&
\omega_2&=(2,3,3,0,-2,-2,-2)^\text{T},\nonumber\\
\alpha_3&=(1,0,0,1,0,-1,-1)^\text{T},&
\omega_3&=(4,4,4,0,-3,-3,-3)^\text{T},\nonumber\\
\alpha_4&=(0,0,0,-2,1,1,1)^\text{T},&
\omega_4&=(2,2,2,-1,-1,-1,-1)^\text{T},\nonumber\\
\alpha_5&=(0,0,0,0,-1,1,0)^\text{T},&
\omega_5&=(3,3,3,0,-3,-2,-2)^\text{T},\nonumber\\
\alpha_6&=(0,0,0,0,0,-1,1)^\text{T},&
\omega_6&=(2,2,2,0,-2,-2,-1)^\text{T},\nonumber\\
\alpha_7&=(0,0,0,0,0,0,-1)^\text{T},&
\omega_7&=(1,1,1,0,-1,-1,-1)^\text{T}.
\end{align}
The nontrivial metric is determined from the duality relation between simple roots and fundamental weights.
In matrix notation it is given by
\begin{align}
G=(\Omega A^\text{T})^{-1},
\end{align}
with $A=(\alpha_1,\alpha_2,\cdots)$ and $\Omega=(\omega_1,\omega_2,\cdots)$.

\section*{Acknowledgements}

We are grateful to H.~Hayashi, N.~Kubo, Y.~Imamura, Y.~L\"u, R.~Ohkawa, M.~Terashita, T.~Tsuda, H.~Tsuhako, especially T.~Nosaka and Y.~Yamada for valuable discussions and comments.
The authors acknowledge the use of ChatGPT (OpenAI) for improving the clarity and readability of this manuscript.
The authors reviewed, edited, and take full responsibility for the final content.
The work of S.M. was supported by JSPS Grant-in-Aid for Scientific Research (C) \#22K03598.

\end{document}